\documentclass[twocolumn,superscriptaddress,aps,prl,amssymb,amsfonts,amsmath]{revtex4-1}
\usepackage{bm}
\usepackage{graphicx}
\usepackage{xcolor}
\newcommand{\rev}[1]{\textcolor{black}{{#1}}}
\definecolor{bleuf}{rgb}{0,0.44,0.72}
\usepackage[unicode=true,pdfusetitle,bookmarks=true,
bookmarksnumbered=false,bookmarksopen=false,breaklinks=false,
pdfborder={0 0 0},backref=false,colorlinks=true,citecolor=bleuf,
linkcolor=bleuf,urlcolor=bleuf]{hyperref}

\begin{document}

\title{{\color{black}Curvature heterogeneities act as singular perturbations to smooth Laplacian fields: a fluid mechanics demonstration}}

\author{St\'ephane Guillet}%
\email{stephane.guillet@ens-lyon.fr}
\author{Benjamin Guiselin}%
\email{benjamin.guiselin@ens-lyon.fr}
\author{Mariem Boughzala}
\author{Vassili Desages}
\author{Denis Bartolo}%
\email{denis.bartolo@ens-lyon.fr}
\affiliation{ENSL, CNRS, Laboratoire de physique, Universit\'e de Lyon, F-69342 Lyon, France}

\date{\today}
\begin{abstract} 
{\color{black}In this Letter, we use a model fluid mechanics experiment to elucidate the impact of curvature heterogeneities on two-dimensional fields deriving from harmonic potential functions. This result is directly relevant to explain the smooth stationary structures in  physical systems as diverse as curved liquid crystal and magnetic films, heat and Ohmic transport in wrinkled two-dimensional materials and flows in confined channels.
 %
Combining microfluidic experiments and theory, we explain how curvature heterogeneities shape confined viscous flows. 
 We show that isotropic bumps induce local  distortions to Darcy's flows whereas anisotropic curvature heterogeneities disturb them  
 algebraically  over system-spanning scales.
 Thanks to an electrostatic analogy, we gain insight into this singular geometric perturbation, and quantitatively explain it using both conformal mapping and numerical simulations. 
 Altogether our findings establish the robustness of our experimental observations and their broad relevance to all Laplacian problems beyond the specifics of our fluid mechanics experiment.} 
\end{abstract}
\maketitle
{\color{black}Trying to elucidate the  dynamics of astronomical objects, Pierre-Simon de Laplace introduced a cornerstone of theoretical physics which we now call Laplace's equation~\cite{Laplace}. 
Since then, far beyond the context of 
celestial mechanics, we now use the solutions of Laplace's equation to model the stationary structures of quantities as diverse as the temperature distribution in materials~\cite{fourier1888theorie}, the concentration of Brownian particles in a solution~\cite{fick}, the electric field away from electric charges~\cite{lagrange1773attraction}, the current distribution in Ohmic conductors~\cite{maxwell}, the wave function of quantum particles~\cite{schrodinger1926quantisierung}, the spin-wave deformations of broken symmetry phases~\cite{Kosterlitz_1973}, and the pressure field of confined fluid flows~\cite{hele1898flow}. 
In particular when an incompressible viscous fluid is driven in the narrow gap separating two large parallel plates (a Hele-Shaw cell), the pressure field $p$ is given by
\begin{equation}
\Delta p=0,
\label{eq:Lapalce}
\end{equation}
and, at scales larger than the gap size, Darcy's law relates the  velocity field to $p$ via~\cite{Guyon_book}: 
\begin{equation}
\mathbf v=-\kappa \bm {\nabla} p.
\label{eq:Darcy}
\end{equation}
The permeability $\kappa$ is a  parameter which embodies the properties of both the liquid and solid walls. 
In the language of Laplacian physics $p$ is  called a harmonic potential function. 
These seemingly mundane relations have elevated the status of the simple Hele-Shaw setup to a powerful experimental tool to investigate Laplacian processes beyond the specifics of fluid mechanics.
Prominent examples include dielectric breakdown~\cite{niemeyer1984fractal}, dendritic growth and transport in disordered media~\cite{Bensimon86,kessler1988,Arneodo89,Barra2001}.

However, aside from rare exceptions,  Darcy's flows and, more broadly, Laplacian phenomena have been mostly studied in flat space.
Very little is known about a basic physics question : How do curvature heterogeneities alter potential flows and other Laplacian processes?
Surprisingly, this fundamental question has been addressed in rather complex situations. From a  non-linear physics perspective, the impact of curvature on Darcy's flows has indeed been limited to interfacial instabilities in model geometries such as cylinders, cones and spheres~\cite{parisio2001saffman, okechi2020stokes, miranda2003geometric, brandao_suppression_2014, brandao_capillary_2017, lee2016fabrication}.
From a condensed matter perspective, most efforts have been devoted to understand how the singular solutions of Laplace's equation, topological defects, couple to curvature in broken-symmetry phases such as superfluids, liquid crystal films and two-dimensional magnetic systems, see Refs.~\cite{turner2010vortices,bowick2009} and references therein.
%
This situation is unsatisfactory not only from a theoretical perspective but also because smooth Laplacian phenomena in curved geometries are realized in numerous experimental situations ranging from Ohmic and heat transport in wrinkled two-dimensional materials~\cite{Ku2020,mohapatra2021thermal}, to breakdown in curved dielectric films~\cite{ji2016boron} and flows in porous media confined between curved fractured rocks~\cite{tsang1998flow}.

\rev{In this Letter, we combine fluid mechanics experiments and theory to reveal and explain how localized curvature heterogeneities generically result in long-ranged perturbations to vector fields that derive from a harmonic potential.
%
For the sake of clarity, we henceforth use the fluid mechanics terminology directly relevant to our experiments.
We first demonstrate that uniform Darcy's flows are merely altered over the footprint of axisymmetric bumps. }
In stark contrast, we  demonstrate that curvature asymmetries  result in algebraic perturbations to both the pressure and velocity fields.
We explain our findings using an electrostatic analogy and conformal mapping arguments.}

\begin{figure*}
    \centering
    \includegraphics[width=0.99\linewidth]{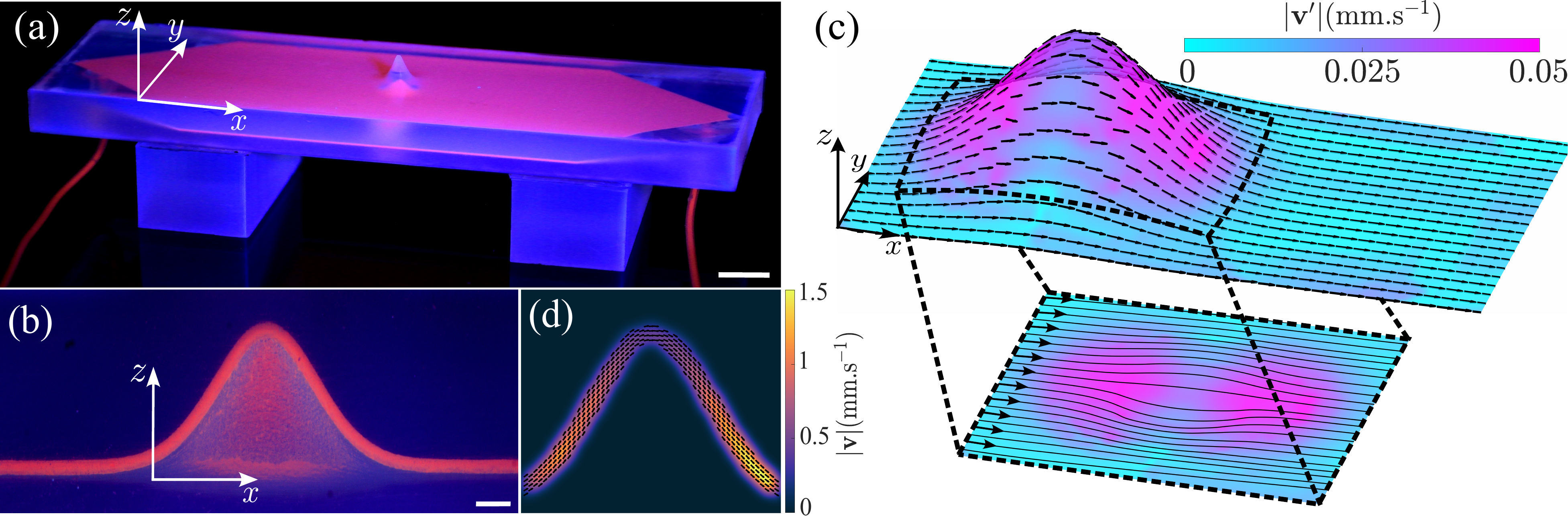}
    \caption{
    (a) Picture of a Hele-Shaw cell made of UV curable resin. It includes a localized curvature heterogeneity in the form of  Gaussian bump located at the center of the channel. The device is filled with UV fabric paint for visualization purpose. 
    ($h_0=3$~mm, $\sigma_x=\sigma_y=1.5$~mm). 
    Scale bar: $8~\rm mm$. 
    (b) 3D print of a half device. This side view of the channel shows that  the gap between the two walls is constant even in the curved regions. 
    Scale bar: $1~\rm mm$. 
    (c) Top: Three-dimensional reconstruction of the velocity field perturbed by a Gaussian bump with $h_0=3$~mm, $\sigma_x=\sigma_y=1.5$~mm.
    The color indicates the magnitude of the flow perturbation $\mathbf{v'}$ .
    Bottom: Corresponding streamlines projected in the $(x,y)$-plane.
    (d) Flow field measured across the channel gap. The black arrows show that the direction of the flow is tangent to the midsurface. The color indicates the magnitude of the velocity field. It has locally a Poiseuille shape. 
    Note that the magnitude of the gap-averaged flow depends on the local geometry of the midsurface. 
    This observation is consistent  with the magnitude of $\mathbf v'$ showed in panel (c). }
    \label{fig:exp_setup}
\end{figure*}
We perform our experiments in three-dimensional (3D) printed microfluidic channels, see Fig.~\ref{fig:exp_setup}(a). 
We make  our channels with a Formlab3 printer and use a transparent photocurable resin (Formlab clear), see also the Supplemental Material (SM)~\cite{SM}. 
In all our experiments the Hele-Shaw cells have a length  $L=120~\rm mm$ and a width $W=38~\rm mm$.
We however perturb the geometry by adding a Gaussian bump at the center of the device as exemplified in Fig.~\ref{fig:exp_setup}(a). 
We stress that the gap of the cells is constant across the whole device, $e=450~$\textmu m, see Fig.~\ref{fig:exp_setup}(b).
%
The geometry of the channels is fully captured by the shape of their midsurface. 
We define it by the height field $h(x,y)=h_0\exp[-x^2/(2\sigma_x^2)-y^2/(2\sigma_y^2)]$, where $\sigma_x,\ \sigma_y$ are the widths of the  bump in the $x$- and $y$-directions respectively. 

We drive the flow with a piezoelectric pressure controller Elveflow OB1 MK4 and image it with  a Hamamatsu ORCA-Quest qCMOS camera mounted on a  Nikon AZ$100$ microscope with a $1.2$ zoom.
We measure the velocity field averaged over the depth of field of our objective in the $(x,y)$ plane. To do so,  we use  a water-glycerol mixture ($20~\rm vol\%$) seeded  with Fluorescent colloidal particles of diameter $4.8$~\textmu m (Thermo Scientific G0500), and perform standard  Particle Imaging Velocimetry~\cite{thielicke2021particle}. 
Knowing $h(x,y)$, we can then project the  velocity  back on the tangent plane and reconstruct the full 3D structure of the flow $\mathbf v(x,y)$ on the curved surface. 
In Fig.~\ref{fig:exp_setup}(c), we can see that the streamlines bend around the bump.

\begin{figure*}
    \centering    \includegraphics[width=\linewidth]{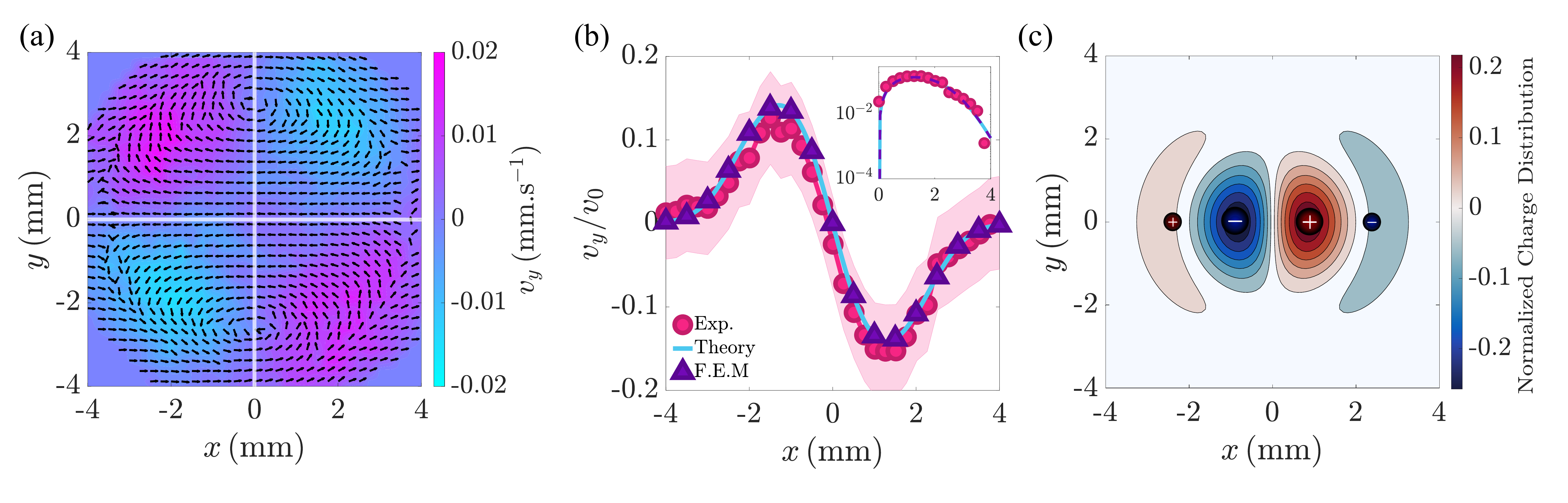}
    \caption{{\bf Isotropic bumps.} (a) Experiments. Planar projection of $\mathbf v'$ in the vicinity of an   axisymmetric Gaussian bump with parameters $h_0=2.25~$mm and $\sigma_x=\sigma_y=1.5$~mm. Away from the bump the fluid flows at a velocity $v_0=120$~\textmu$\rm m.s^{-1}$ along the $x$-direction. 
    The velocity field has the same angular symmetry as the electric field induced by a charge dipole antiparallel to the unperturbed flow. 
    (b) Experiments, theory and simulations. $y$-component of the velocity perturbation $v_y'$ for an axisymmetric Gaussian bump ($h_0=3~$mm,  $\sigma=\sigma_x=\sigma_y=1.5$~mm). $v_y'$ is plotted as a function of the coordinate $x$ in the direction of the unperturbed flow. 
    The velocity is measured  at $y/\sigma = 0.78 \pm 0.09$.
    The theoretical results derived from conformal theory in an infinite channel, and from FEM simulations in a finite channel are in excellent agreement with our measurements.
    Inset: log-lin plot of the same data. 
    (c) Theory. Color map of the equivalent normalized charge distribution $\kappa\sigma^3\lambda/(v_0h_0^2)$. 
    It roughly corresponds to the  superposition of two antiparallel dipoles that exactly cancel out.}
    \label{fig:data_sym}
\end{figure*}

To better quantify these flow perturbations, we define  $\mathbf v'=\mathbf v- \mathbf v_0$, where, $\mathbf v_0=v_0\hat{\mathbf x}$ is the uniform flow measured at large distance from the bump.  
Fig.~\ref{fig:data_sym}(a) shows that $\mathbf v'$ has a clear dipolar symmetry, akin to the flow around a fixed obstacle in a flat channel~\cite{Guyon_book}. 
Given this symmetry, and the Laplacian nature of the hydrodynamic problem, one would expect $\vert\mathbf{v}'\vert$ to decay algebraically with $\rho=\sqrt{x^2+y^2}$, the distance to the apex of the bump, ($\vert\mathbf{v}'\vert\propto 1/\rho^2$, like the electric field induced by a charge dipole in two dimensions)~\cite{Beatus2006,Cui2004}, see also SM~\cite{SM}.  
This prediction is however at odds with our measurements. 
The flow perturbations are exponentially localized in space within the footprint of the bump. 
This  atypically fast decay is better seen in Fig.~\ref{fig:data_sym}(b), where we plot the $y$-component of $\mathbf v'$ as a function of $x$. 

This counter intuitive effect  begs for a theoretical explanation.  
To address this question analytically and numerically, we first need to write the covariant generalizations of Darcy's law and mass conservation on a curved surface. 
They take the compact form 
\begin{align}
v^\alpha=-\kappa g^{\alpha\beta}\partial_\beta p,
\label{eqn:Darcy_curved}
\\
 \frac{1}{\sqrt{g}} \partial_\alpha\left(\sqrt{g}v^\alpha\right)=0.
 \label{eqn:Mass_curved}
\end{align}
 $v^\alpha$ ($\alpha=x, \ y$) is the $\alpha$-component of the velocity field in the local basis $(\mathbf{e_x},\ \mathbf{e_y})$ of the tangent plane to the midsurface, $\mathbf{e_x}=\mathbf{\hat{x}}+\mathbf{\hat{z}}\,\partial_x h$, $\mathbf{e_y}=\mathbf{\hat{y}}+\mathbf{\hat{z}}\,\partial_y h$, and  $g_{\alpha\beta}=\delta_{\alpha\beta}+\partial_\alpha h\,\partial_\beta h$ is the associated metric.
 Combining these two equations we find that the pressure field $p$ obeys the Laplace-Beltrami equation:  $(1/\sqrt{g})\partial_\alpha(\sqrt{g}g^{\alpha\beta}\partial_\beta p)=0$, see SM for a detailed derivation of the above equations~\cite{SM}. Equations~\eqref{eqn:Darcy_curved}  and \eqref{eqn:Mass_curved} tell us that any local change in the metric should alter Darcy's flows.

When the geometry of the channel is modified by an axisymmetric bump $h(\rho)$,  we  can compute the pressure and flow fields by  taking advantage of the conformal invariance of the Laplace-Beltrami operator~\cite{bazant2004conformal}. We consider a single bump on an infinite surface and a uniform flow away from the bump. 
We can then flatten the surface  using a global conformal map described by the conformal  factor $\Omega(\rho)$, solve  Laplace's equation in the plane, and finally apply the inverse transform to compute the pressure and velocity fields on the bumpy surface, see SM~\cite{SM}. 
Regardless of the specific shape of the bump, the  conformal factor is given by~\cite{vitelli2004Defect}:
\begin{equation}
\Omega(\rho)=\exp\left[\int_\rho^{+\infty}\frac{\mathrm{d}\varrho}{\varrho}\left(\sqrt{1+h'(\varrho)^2}-1\right)\right].
\end{equation}
Fig.~\ref{fig:data_sym}(b) shows that our analytical solution is in excellent agreement with our experimental findings.
This agreement  establishes that the measured deviation to a uniform flow field  originates  from curvature heterogeneities. 
We also note that conformal invariance readily informs us on the range of the flow perturbations.
Isotropic bumps are conformally flat, in other words we can deform them into a planar surface by applying a {\em global} map that solely involves {\em local} dilations of the curved metric. 
Away from the bump, the local dilation factor becomes vanishingly small and therefore the solution of  Eqs.~\eqref{eqn:Darcy_curved} and \eqref{eqn:Mass_curved} reduces to the uniform flow of a flat Hele-Shaw cell. 
The decay of $\vert \mathbf v'\vert$ is therefore  set by the decay of $\Omega(\rho)-1$: for Gaussian bumps $\mathbf v'$ is exponentially localized in space. 
 
To understand deeper the singular range of the flows around axisymmetric curvature heterogeneities, we now take advantage of an electrostatic analogy. 
In the limit of small aspect ratio $a=h_0/\sigma\ll 1$ ($\sigma = \sigma_x=\sigma_y$), we expand Eqs.~\eqref{eqn:Darcy_curved} and \eqref{eqn:Mass_curved} to second order in $a$ and find that the lowest-order correction to the pressure field satisfies Poisson's equation:
\begin{equation}
    \Delta p^{(2)}=-\kappa^{-1}(\mathbf v_0\cdot\bm{\nabla} h)\Delta h,
    \label{eq:p2}
\end{equation}
with $p^{(2)}(x,y)\to0$ when $x,y\to\infty$ (see also SM~\cite{SM} for a detailed derivation). 
Computing the first correction to the pressure field is thus equivalent to finding the electric potential induced by a charge distribution $\lambda = \kappa^{-1}(\mathbf v_0\cdot\bm{\nabla} h)\Delta h$. 
To gain some intuition about the form of the pressure fluctuations, we plot the charge distribution $\lambda(x,y)$ in Fig.~\ref{fig:data_sym}(c).
In the far-field limit, the equivalent charge distribution can be seen  as the sum of two dipoles pointing in opposite directions as the signs of the local slope $(\mathbf v_0\cdot\bm{\nabla} h)=v_0\partial_x h$, and mean curvature $\Delta h$, change once and twice respectively across the bump. 
The first dipole is  formed by strong charges  separated by a small distance, while the second dipole is formed by weaker charges separated by a larger distance, see Fig.~\ref{fig:data_sym}(c).
\begin{figure*}
    \centering
    \includegraphics[width=\linewidth]{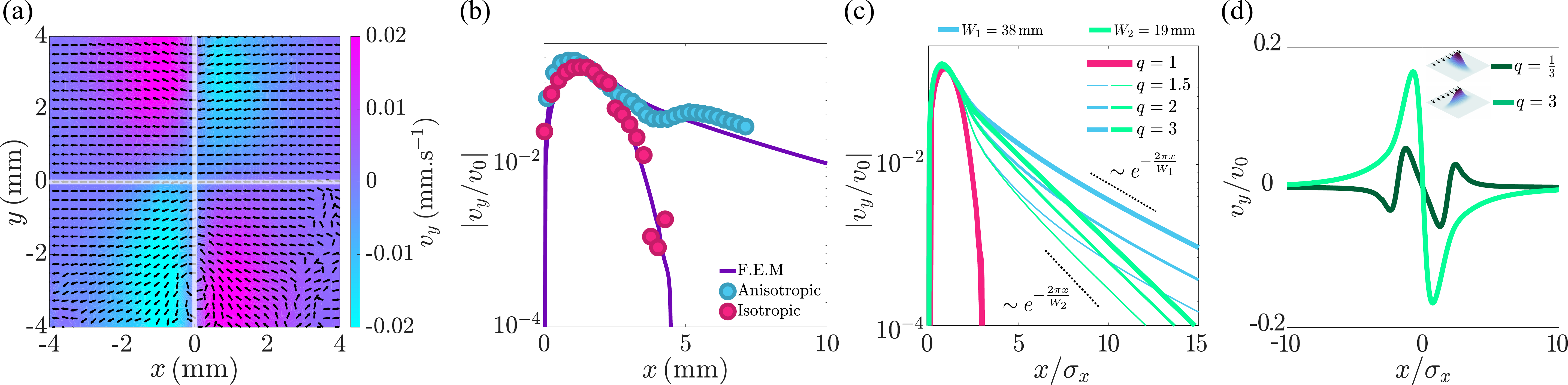}
    \caption{
    {\bf Anisotropic bumps.}
    (a) Experiments.  Planar projection of $\mathbf v'$ in the vicinity of an anisotropic Gaussian bump with parameters $h_0=3.5$mm, $\sigma_x=1.5$~mm and $\sigma_y=4.5$~mm.
    The streamlines of the velocity field retain a dipolar symmetry. 
    (b) Experiments and simulations. Log-lin plot of the perturbation to the mean flow for an isotropic Gaussian bump with parameters $h_0=3~$mm, $\sigma_x=\sigma_y=1.5$~mm and for an anisotropic bump with parameters $h_0=3.5$mm, $\sigma_x=1.5$~mm, $\sigma_y=4.5$~mm as a function of the coordinate $x$ in the direction of the unperturbed flow for $y/\sigma_y = 0.78 \pm 0.09$. 
    We plot both the experimental and FEM results.
    (c) FEM simulations. Log-lin plot of the perturbation to the mean flow for four diffent anisotropic bumps in channels of two different widths $W$. 
    The decay of $\vert\mathbf{v'}\vert$ is set by the shape of the isotropic  bump, by contrast $\vert\mathbf{v'}\vert$ decays much slower for anisotropic bumps. In the far-field limit, the algebraic decay is exponentially screened over a distance set by the channel width.
    (d) FEM simulations. Lin-lin plot of the perturbation to the mean flow for two anisotropic bumps pointing in orthogonal directions, see sketches, opposite anisotropies obtained from FEM simulations.
    The sign of the far-field perturbation changes when the orientation of the main axis of the bump exceeds $45^{\circ}$.}
    \label{fig:data_asym}
\end{figure*}
For any axisymmetric bump, the two dipoles cancel out exactly. To see this, we can compute the net dipole associated with the charge distribution $\mathbf{P}=\iint\mathrm{d}x\,\mathrm{d}y\,\lambda(x,y)\left(x\,\mathbf{\hat{x}}+y\,\mathbf{\hat{y}}\right)$, and find
\begin{equation}
\mathbf P=\frac{\pi v_0h_0^2}{4\kappa}\left(\frac{\sigma_x}{\sigma_y}-\frac{\sigma_y}{\sigma_x}\right)\mathbf{\hat{x}}.
\label{eq:dipole}
\end{equation}
$\mathbf P$ vanishes when $\sigma_x=\sigma_y$, 
as well as all higher-order multipoles as shown in SM~\cite{SM}. 
Therefore, despite the existence of a non-trivial charge distribution, the far-field correction to  the pressure around an isotropic bump cannot be captured by any multipolar expansion.
The fluctuations of the flow field require some more attention. $\mathbf v'(\mathbf r)$ has indeed both a kinematic and a dynamic origin which are both impacted by curvature. At a perturbative level $\mathbf v'=v_0 g^{(2)}: \mathbf{e_x}-\kappa\left[\partial_x p^{(2)}\mathbf{e_x}+\partial_yp^{(2)}\mathbf{e_y}\right]$, where $g^{(2)}$ is the second-order correction to the inverse metric. The first term is a mere kinematic correction that stems from the projection of the unperturbed flow field on the tangent plane. 
The second term is specific to Darcy's flows and may carry non-local perturbations to the velocity field. It is analogous to the electric field induced by the charge distribution $\lambda$. When $\sigma_x=\sigma_y$ both terms vanish in the far-field limit:
all curvature-induced flows are screened past the footprint of the bump.

{\color{black} We note in passing that the agreement between our experimental observations and our theory justifies the relevance of our two-dimensional Laplacian theory despite the weak scale separation between the gap size and the spatial extent of the bump. With hindsight this agreement is however not surprising as typical Brickman's corrections would result in corrections to the velocity field of the order of $[e/(\pi\sigma_x)]^2\approx 10^{-2}$~\cite{zeng2003brinkman}.}

We now move to  our second central result, as clearly seen in Eq.~\eqref{eq:dipole},  our results heavily rely on rotational symmetry. 
We therefore need to address the impact of curvature anisotropy, which would exist in any natural setting. 
At a perturbative level, the electrostatic analogy tells us that the flow perturbations should be long-ranged regardless of the specific functional form of $h$,   $\vert\mathbf{v'}\vert$ should decay as $1/\rho^2$, see Eq.~\eqref{eq:dipole}.
To assess whether this prediction holds beyond perturbation theory,
it would be tempting to use the same theoretical tools as above. 
Namely to look for a global conformal transformation that would map anisotropic bumps onto planar domains~\cite{blanc1941filetype}.
We indeed know that simple conformal maps transform anisotropic domains into isotropic ones, such as the celebrated Joukowski transform  in the context of fluid mechanics. 
These maps should however cause  long-range correlations in the $\mathbf v'(x,y)$ field, as they typically involve inversions, which are {\em non-local} transformations. 
There is therefore no reason to expect any geometrical screening of the curvature-induced perturbations.

To further confirm our reasoning, we first conduct experiments in channels deformed by Gaussian bumps with $\sigma_y>\sigma_x$ and measure $\mathbf v'$, see Fig.~\ref{fig:data_asym}.
We find that the angular symmetry of the perturbation remains 
mostly dipolar, see Fig.~\ref{fig:data_asym}(a).
But, Fig.~\ref{fig:data_asym}(b) shows that the perturbation induced by anisotropic bumps extends as expected over much larger distances. 

However,  unlike our theoretical prediction, $\vert\mathbf v'\vert$ still decays exponentially. 
The finite width of our channels  explains this much simpler screening effect.
Using again our electrostatic analogy, the two side walls of the channel acts as two conductors which  screen the electric potential induced by a point dipole over a distance $W/(2\pi)$, see also SM~\cite{SM}. 
To quantitatively check our reasoning, we perform FEM simulations of Eqs.~\eqref{eqn:Darcy_curved} and~\eqref{eqn:Mass_curved}, see SM~\cite{SM}.
Fig.~\ref{fig:data_asym}(b) shows an excellent agreement with our measurements, and Fig.~\ref{fig:data_asym}(c) confirms that the far-field decay of the flow is not set by the bump geometry but by the channel width only.

As a last comment, we note that our electrostatic analogy  informs us on the sign of the far-field perturbation as well.  As $\mathbf{P}\propto\mathbf{\hat x}$ when $\sigma_x>\sigma_y$, and $\mathbf{P}\propto-\mathbf{\hat x}$ otherwise, the far-field perturbation should be positive (resp. negative) when the long axis of the bump makes a smaller angle with the $y$-axis (resp. $x$-axis).
Our FEM simulations again confirm that this last prediction holds even in the limit of high bumps, see Fig.~\ref{fig:data_asym}(d).

{\color{black}To conclude, we have shown that unlike Stokes flows~\cite{Davidovitch2022}, static curvature heterogeneities generically deform the streamlines of  Darcy's flows. 
Combining microfluidic experiments and theory we have revealed that curvature anisotropy acts as a singular perturbation to potential flows. 
Using a robust electrostatic analogy we have explained why the flow distortions induced by isotropic bumps are screened while vanishingly-small curvature asymmetries could bend the streamlines  over system-spanning scales.
We hope that our findings will stimulate a deeper investigation of the role of curvature heterogeneities on a broader class of Laplacian phenomena ranging from superfluid film flows to transport in fractured rocks and Ohmic transport in wrinkled two-dimensional conductors. We also expect our results to be relevant to non-Laplacian problems that can be solved using conformal mappings, such as advection-diffusion in potential flows, electrochemical transport~\cite{bazant2004conformal}, and possibly  stress propagation and adhesion pattern formation in elastic films~\cite{mitchell2017fracture,Hure2011}.}

\begin{acknowledgments}
We thank Eran Sharon and Benny Davidovitch for insightful comments and suggestions, and Alexis Poncet for a careful reading of the manuscript. S.~G. and B.~G. have equally contributed to this work.
\end{acknowledgments}

\bibliography{./biblio.bib}
\end{document}


\title{Supplemental Material: \\
Flowing in Curved Space: Curvature Heterogeneity and Singular Perturbation to Darcy's Flows}
\author{St\'ephane Guillet}%
\email{stephane.guillet@ens-lyon.fr}
\author{Benjamin Guiselin}%
\email{benjamin.guiselin@ens-lyon.fr}
\author{Mariem Boughzala}
\author{Vassili Desages}
\author{Denis Bartolo}%
\email{denis.bartolo@ens-lyon.fr}
\affiliation{ENSL, CNRS, Laboratoire de physique, Universit\'e de Lyon, F-69342 Lyon, France}
\maketitle

\setcounter{equation}{0}
\setcounter{figure}{0}
\setcounter{table}{0}
\setcounter{page}{1}
\renewcommand{\thefigure}{S\arabic{figure}}
\renewcommand{\theequation}{$\mathcal{S}$\arabic{equation}}
\renewcommand{\bibnumfmt}[1]{[S#1]}
\renewcommand{\citenumfont}[1]{S#1}
\renewcommand\thesubsection{\arabic{subsection}}
\renewcommand\thesubsubsection{\arabic{subsection}.\arabic{subsubsection}  }

\section{Experimental methods}
\subsection{3D printed curved Hele-Shaw cell}
We first detail  the procedure to render the three-dimensional (3D) models of the curved channels shown in Fig.~1.  
We generate a three-dimensional Gaussian surface as defined  in the main text  using the Mathematica software. 
In practice, for isotropic bumps, we set the height function to zero for $\sqrt{x^2+y^2}\geq \sigma$ ($\sigma=\sigma_x=\sigma_y)$, and 
for $\abs{x}\geq 3\sigma_x$, or $\abs{y}\geq3\sigma_y$, in the case of anisotropic bumps. 
We then  numerically translate the surface over  a distance $\pm \,e/2$ along the direction of the  local normal vector 
\begin{equation}
    \mathbf{n}=\frac{1}{\sqrt{1+(\partial_x h)^2+(\partial_y h)^2}}\begin{pmatrix}
        -\partial_x h\\
        -\partial_y h\\
        1
    \end{pmatrix}
\end{equation}
in $(x,\ y,\ z)$ Cartesian coordinates. We then tesselate the two surfaces and embed the resulting mesh into a STL Computer-Aided Design (CAD) file.
The resulting 3D object defines the boundaries of the channel of thickness $e$. 
We complete the  full channel geometry  by importing the .STL file to the CAD software Fusion$360$ where we perform additional basic operations of extrusion and subtraction.

The 3D model of the channel is exported to a Formlabs SLA 3D printer (Form$3$) using the Formlabs print preparation software (Preform). 
 To improve the resolution of the 3D prints, the direction transverse to the main channel coincides with the direction along which the laser spot is optically translated (as opposed to mechanically).
 The mechanical displacement steps are set to their minimal value, namely, 25~\textmu m. 
 The printing time is about 9~hours.

The resulting device is washed for $10$ minutes in Isopropyl alcohol (IPA) using the  Formlabs washing machine (FormWash), and  sonicated for $5$ minutes in an IPA bath. 
We finally flush the remaining resin out of the channel using pressurized nitrogen gas. 
These three steps are repeated until the inner channel is totally free from any residual uncured resin.   

The channels are made of the translucent Formlabs Clear resin to allow optical observations.
However, the outer surfaces of the 3D printed channels have a finite roughness that makes them opaque.
To improve the optical transparence, we coat the outer surfaces with a thin layer of Clear resin on top which we  cure for $5$ minutes at $60^\circ$C thanks to a UV lamp of wavelength 405~nm (FormCure).
This procedure suppress most of the unwanted light scattering and makes optical observations of the flow possible.

We inject the fluids using ETFE (Tefzel) tubes of inner diameter $1/16"$, which we  seal with epoxy resin at the inlet and outlet of the channel. 

\subsection{Characteristics of the fluid}
We match the refractive index of the fluids and of the cured resin to further improve the optics. 
To do so, we use a suspension (0.1~vol\%) of PIV tracers in a water-glycerol solution (20~\rm vol\%). 
The tracers are polysterene colloids of diameter $4.8~$\textmu m  (Thermo Scientific G0500)
 We control the flows thanks to an Elveflow pressure controller OB1 MK4, and the flow rate is retroactively adjusted to $60$~\textmu L/min using a  Microfluidic flow sensor (Elveflow). 

We wait a few minutes to reach a steady state and start our measurements.

\subsection{PIV measurements}

The channel is observed with Nikon AZ$100$ microscope with $\times 1$ and $\times 2$ magnification objectives (Nikon AZ plan Fluor). 
We record images of the flow using an ORCA-Quest qCMOS camera (C15550-20UP)  mounted on the microscope thanks to a DSC port which introduces an extra $\times 0.6$ magnification. 
The  field of view is typically a rectangle of size 16~mm$ \times$ 9~mm (corresponding to 4096~px $\times$ 2304~px) in the $xy$-plane. 
The colloidal particles are imaged with a fluorescent light of wavelength 468~nm (Nikon Intensilight C-HGFI).
We record $300$ pictures taken with an exposure time of 100~ms and a frame rate of 10~fps.

\begin{figure}
    \centering
    \includegraphics[width=0.99\linewidth]{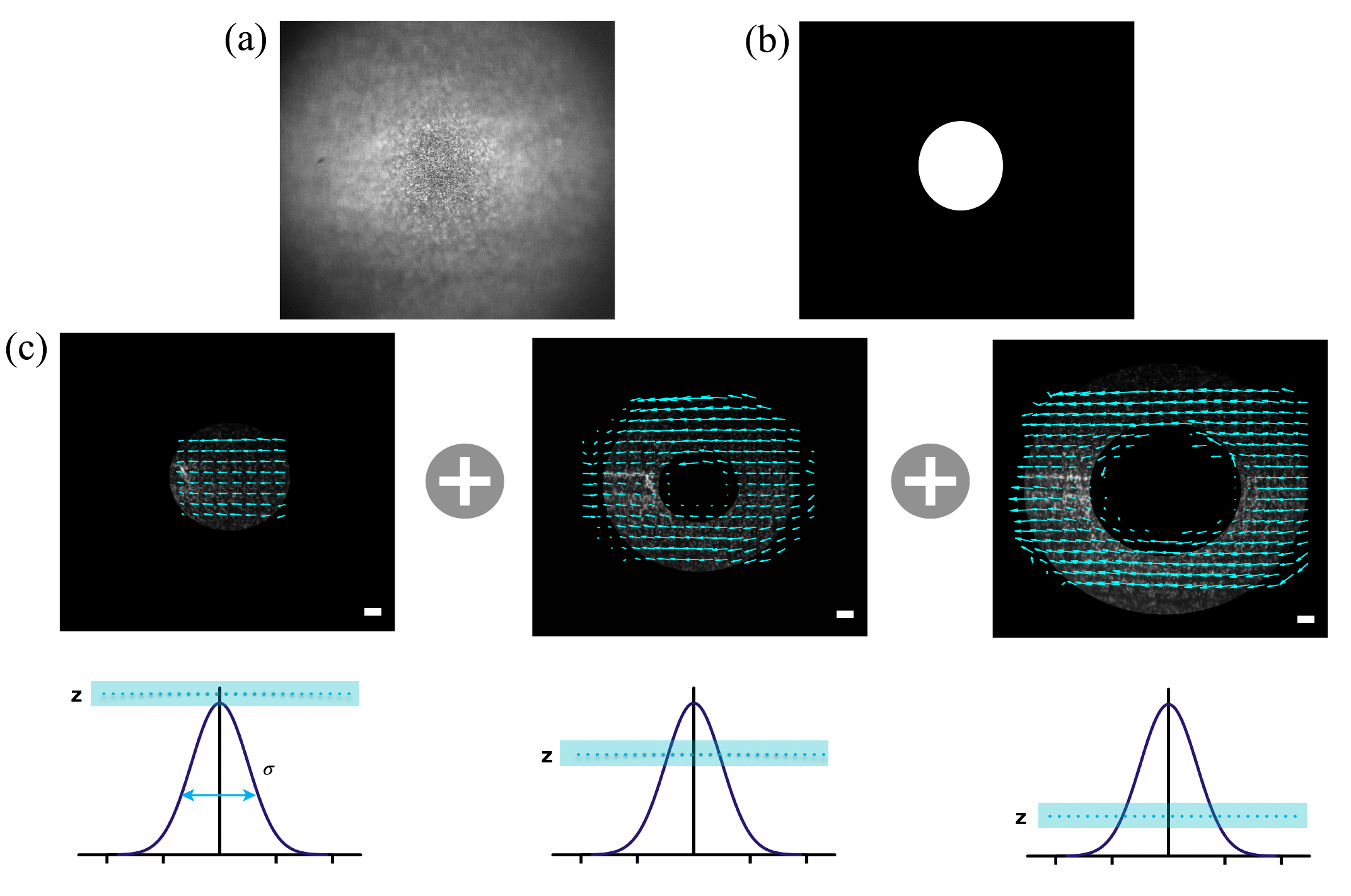}
    \caption{
    Schematic representation of the Particle Image Velocimetry (PIV) protocol. 
    (a) Picture of a water-glycerol mixture seeded with fluorescent colloïdal particles in the vicinity of a Gaussian bump.
    (b)   Illustration of a typical Region of Interest (ROI) captured at the apex of the Gaussian bump. The white disk corresponds to the region on focus.
    (c) PIV measurements  at various focal planes. The mean flow is from right to left. 
    Scale bar: $0.1~\rm mm$. 
    }
    \label{fig:SIPIV}
\end{figure}

Figure~\ref{fig:SIPIV} explains schematically our PIV method. We detail it below.
The  raw images are preprocessed prior to being analyzed with the PIV algorithm. 
In practice, a circular mask is applied to remove blurred areas which are outside the depth of field of the microscope. We then apply a constant threshold to remove the unwanted background noise, and  optimize the contrast by multiplying the intensity by a constant factor of 2.5. Finally we apply a Gaussian blur with a standard deviation of $2$~px to smooth the intensity signal.

We are then equipped to systematically perform PIV measurements using the Pivlab Matlab package~\cite{thielicke2021particle}.
We reconstruct the velocity field in the field of view, which coincides with the $(x, y)$-plane, see Fig.~1 in the main text. 
We compute the velocity field using a time interval of 3 frames, which corresponds to an optimal displacement of the colloids of about one diameter. 
We use three PIV passes performed in PIV windows of size 512, 256 and 128 pixels separated by a half window size.
Since the flow is stationary, we finally compute the mean velocity at a given position by averaging the measured velocity over all frames. 

In a typical data acquisition, the apex of the bump is approximately centered in the field of view, and the focus is first made at the top of the bump. 
After taking the $300$ pictures, we manually change the focus to a lower plane of the bump. 
To scan the whole bump, we typically reiterate this procedure $7$ to $10$ times (for bumps of height $h_0=3.5$~mm).
We then translate the deck-plate by a distance of $1000$~px (about 2~mm) along the $x$-axis or $y$-axis. 
We  reconstruct the whole flow map  by stitching the velocity sub-maps and averaging the field in the overlapping regions.

To measure the component of the velocity field in the $z$-direction (optical axis direction), we make the following  observation: the velocity field is tangent to the midsurface of the bump. Therefore $v_z=v_x\partial_xh+v_y\partial_yh$. 
To use this formula, we however need to locate the apex of the bump accurately to define the origin of our frame.
We do this by analyzing the variations  of $v_y(x,y)$ along several horizontal and  vertical cross sections and note that   $v_y(0,0)$ must vanish at the apex given the mirror symmetry of our devices. 

We end this section by mentioning that the velocity profiles presented in the main text are smoothed by performing an average over the nearest neighboring data points. 

\section{Theory and numerical resolution}
\subsection{Outline}
This section is organized as follows. 
We first explain how to solve exactly the flow induced by an isotropic bump using conformal mapping techniques. In other words we solve Darcy's law and mass conservation equation [Eqs.~(1) and (2) in the main text] which reduce to the Laplace-Beltrami problem 
\begin{equation}
\frac{1}{\sqrt{g}}\partial_\alpha\left(\sqrt{g}g^{\alpha\beta}\partial_\beta p\right)=\Delta_\mathrm{LP}p=0,
    \label{eqn:laplace_curved}
\end{equation}
where $\Delta_\mathrm{LP}$ is the Laplace-Beltrami operator.

We then take a step back and give a more physical insight into this solution using an analogy with electrostatics. 
In particular we show that we can relate the perturbation to the flow field to the electric field induced by a charge distribution defined by the geometry of the bump. 
This analogy explains why isotropic bumps result in perturbations localized over their footprint whereas anisotropic curvature heterogeneities cause algebraic perturbations to Darcy's flows.

In the limit of small bumps, we can also account for the finite width of the channel using the same theoretical framework and show that the flow perturbations are exponentially screened by confining walls.

Equipped with all these predictions, we test them against numerical resolutions of Eq.~\eqref{eqn:laplace_curved} done using a Finite Element Method solver.

Lastly, we provide a thorough justification of the covariant version of Darcy's law starting from the most general description of Stokes flows confined in channels having a curved geometry and a finite thickness.

\subsection{Exact solution for an axisymmetric bump from conformal transformations}
\label{sec:conformal}

In this section, we solve Eq.~\eqref{eqn:laplace_curved} without any assumption about the magnitude of the curvature heterogeneities.
In the case of isotropic bumps, this is achieved thanks to a conformal mapping of the surface onto the flat two-dimensional Euclidean plane. 
Given this symmetry, we consider a height function $h(\rho,\phi)=h(\rho)$ where $\rho$ and $\phi$ are the polar  coordinates.
The metric of the surface  is given by
\begin{equation}
    g_{\alpha\beta}=\begin{pmatrix}
    1+h'(\rho)^2&0\\0&\rho^2
    \end{pmatrix},
\end{equation}
and Eq.~(\ref{eqn:laplace_curved}) then reads
\begin{equation}
    \displaystyle \frac{\rho}{\sqrt{1+h'(\rho)^2}}\partial_\rho\left(\frac{\rho}{\sqrt{1+h'(\rho)^2}}\partial_\rho p\right)+\partial_{\phi\phi}p=0.
\end{equation}
We now use conformal invariance. 
The two-dimensional curved surface defined by the metric $g_{\alpha\beta}$ can be conformally flattened with a single map. 
Simply put, there exists a set of coordinates $\vec{y}$=($\mathcal{R}$, $\phi$) such that 
\begin{equation}
g_{\alpha\beta}\d x^\alpha\d x^\beta=\Omega(\rho)^2\eta_{\alpha\beta}\d y^\alpha\d y^\beta,
\label{eqn:metrics}
\end{equation}
with $\vec{x}$=($\rho$, $\phi$), and $\eta_{\alpha\beta}$ the flat-space metric in polar coordinates.
$\Omega(\rho)$ is the so-called conformal factor which only depends on $\rho$ because the surface is isotropic. 
This conformal map corresponds to a local dilation of the surface.
The amplitude of the local dilation is given by the conformal factor deduced from Eq.~\eqref{eqn:metrics}:
\begin{equation}
    \d\mathcal{R}^2=\frac{1+h'(\rho)^2}{\Omega(\rho)^2}\d\rho^2,\quad\mathcal{R}^2=\left(\frac{\rho}{\Omega(\rho)}\right)^2.
\end{equation}
By combining the  above equations, we find~\cite{vitelli2004Defect}
\begin{equation}
    \frac{\d\mathcal{R}}{\mathcal{R}}=\pm\frac{\sqrt{1+h'(\rho)^2}}{\rho}\d\rho,
\end{equation}
where we keep the solution with a positive sign, because we want $\mathcal{R}$ to increase when $\rho$ grows. 
This equation is straightforward to solve and leads to
\begin{equation}
\begin{aligned}
    \mathcal{R}(\rho)&=\mathcal{R}_0\exp\left[\int_{\rho_0}^\rho\frac{\d\varrho}{\varrho}\sqrt{1+h'(\varrho)^2}\right],\\
    &=\frac{\mathcal{R}_0}{\rho_0}\rho\exp\left[\int_{\rho_0}^\rho\frac{\d\varrho}{\varrho}\left(\sqrt{1+h'(\varrho)^2}-1\right)\right],
    \end{aligned}
\end{equation}
where $\rho_0$ and $\mathcal{R}_0$ are two free parameters. 
In the limit $\rho\to+\infty$ (or outside the domain where the height function is non-zero), the surface is flat [$h'(\rho)\to0$].
We can then impose that the two sets of coordinates coincide, namely that $\mathcal{R}(\rho)\sim\rho$.
This condition sets $\mathcal{R}_0=\rho_0$, (taking $\rho_0\to+\infty$), and we eventually find
\begin{equation}
    \mathcal{R}(\rho)=\rho\exp\left[-\int_\rho^{+\infty}\frac{\d\varrho}{\varrho}\left(\sqrt{1+h'(\varrho)^2}-1\right)\right]=\frac{\rho}{ \Omega(\rho)}.
\end{equation}
This expression  allows us to determine the conformal factor $\Omega(\rho)$, given by Eq.~(3) in the main text.

The significant advantage offered by of the conformal transformation is that the Laplace-Beltrami equation merely reduces to Laplace's equation in the conformally flat space. 
We can solve it using Cartesian coordinates, $\mathcal{X}=\mathcal{R}\cos(\phi)$ and $\mathcal{Y}=\mathcal{R}\sin(\phi)$. 
The full pressure problem then takes the form:
\begin{equation}
\begin{cases}
    \partial_{\mathcal{X}\mathcal{X}}p+\partial_{\mathcal{Y}\mathcal{Y}}p=0,\\
    \partial_{\mathcal{X}}p(-L/2,\mathcal{Y})=-v_0/\kappa,\\
    p(L/2,\mathcal{Y})=0,\\
     \partial_{\mathcal{Y}}p(\mathcal{X},\pm\,W/2)=0.
    \end{cases}
    \label{eqn:laplace_conformal_axisym}
\end{equation}
We note that the boundary conditions we chose require  the height function to have a finite support, or to  decay exponentially. In other words we assume the surface to have a vanishing curvature at the boundaries (rectangle of length $L$ and width $W$). 
In the flattened space, the pressure field is the well-known affine function in the direction of the velocity:
\begin{equation}
    p(\mathcal{X},\mathcal{Y})=\frac{v_0}{\kappa}\left(\frac{L}{2}-\mathcal{X}\right).
\end{equation}
Given this solution, we can now come back to the original curved space, and provide an exact expression for the pressure field
\begin{equation}
    p(\rho,\phi)=\frac{v_0}{\kappa}\left[\frac{L}{2}-\frac{\rho}{\Omega(\rho)}\cos(\phi)\right]=\frac{v_0}{\kappa}\left\{\frac{L}{2}-\rho\cos(\phi)\exp\left[-\int_\rho^{+\infty}\frac{\d\varrho}{\varrho}\left(\sqrt{1+h'(\varrho)^2}-1\right)\right]\right\}.
\end{equation}
We also find  the velocity field using Eq.~(1) in the main text: $v^\alpha=-\kappa g^{\alpha\beta} \partial_\beta p$.
In Fig.~2(b) in the main text we plot this solution for the specific case of a Gaussian bump parametrized by
\begin{equation}
h(\rho)=\begin{cases}
    h_0 e^{-\rho^2/(2\sigma^2)}\ \text{if}\ \rho\leq 3\sigma,\\
    0\ \text{otherwise}.
\end{cases}
\end{equation}

\subsection{Gaining some physical insights: an electrostatic analogy}

\subsubsection{Perturbative analysis}
To gain more physical insight into the coupling between flows and curvature, it is worth introducing a perturbative analysis of the hydrodynamic problem.
We consider a surface $z=h(x,y)$ parametrized in the Monge gauge where $(x,\ y)$ are Cartesian coordinates. 
We denote $\sigma_x$ and  $\sigma_y$ the typical widths of the the bump in the $x$ and $y$-directions, and $h_0=h(0,0)$ the maximum of $h$. 

We focus here on the limit of smooth surface heterogeneities, {\it i.e.} on the limit where the bump aspect ratio $a=h_0/\sigma_x$ is small.
The geometry of the surface is characterized by its $2\times 2$ metric tensor
\begin{equation}
g_{\alpha\beta}=\delta_{\alpha\beta}+\partial_\alpha h\, \partial_\beta h.
\end{equation}
We introduce the rescaled quantities $\tilde x=x/\sigma_x$, $\tilde y = y/\sigma_x$, $\tilde h(\tilde x,\tilde y)=h(x,y)/(a\sigma_x) $ and   the anisotropy factor $q=\sigma_y/\sigma_x$.
In the specific case of a Gaussian bump  $\tilde h(\tilde x, \tilde y)=e^{-\tilde x^2/2-\tilde y^2/(2q^2)}$.

We can now perform a Taylor expansion of all the geometric quantities in  powers of $a$. At leading order, we find
\begin{equation}
\begin{cases}
g_{\alpha\beta}=\delta_{\alpha\beta}+a^2\tilde\partial_\alpha\tilde h\,\tilde\partial_\beta\tilde h,\\
g^{\alpha\beta}=\delta^{\alpha\beta}-a^2\delta^{\alpha\gamma}\delta^{\beta\epsilon}\tilde\partial_\gamma \tilde h\, \tilde\partial_\epsilon \tilde h+O(a^4),\\
\sqrt{g}=1+\dfrac{1}{2}a^2(\mathbf{\tilde \nabla}\tilde h)^2+O(a^4),
\end{cases}
\end{equation}
with $\tilde \partial_x$ (resp. $\tilde \partial_y$) the partial derivative with respect to coordinate $\tilde x$ (resp. $\tilde y$), and $\mathbf{\tilde\nabla}=\mathbf{\hat x}\,\tilde\partial_x+\mathbf{\hat y}\,\tilde\partial_y$. 
Equation~\eqref{eqn:laplace_curved} then reduces to 
\begin{equation}
\begin{cases}
    \tilde \Delta p^{(0)}=0,\\
    \tilde \Delta p^{(2)}=\tilde\Delta\tilde h\left(\mathbf{\tilde\nabla}\tilde h\cdot\mathbf{\tilde\nabla}p^{(0)}\right)+\delta^{\alpha\gamma}\delta^{\beta\epsilon}\tilde\partial_\gamma\tilde h\tilde\partial_\epsilon\tilde h\tilde\partial_{\alpha\beta}p^{(0)},
    \end{cases}
\end{equation}
 where we have expanded the pressure field up to second order in $a$:
\begin{equation}
p(x,y)=p^{(0)}(\tilde x,\tilde y)+a^2p^{(2)}(\tilde x,\tilde y)+O(a^4).
\end{equation}
We are therefore left with a Poisson problem to solve (for $p^{(2)}$). To solve this equation we need to prescribe a set of boundary conditions. In this work we consider curvature heterogeneities that are localized in space. In more quantitative terms, we consider $h$ functions which are at most exponentially localized in space.
In the limit where $\sigma_{x,y}\ll W$, we can set the boundary conditions
\begin{equation}
    \partial_x p(-L/2,y)=-v_0/\kappa, \quad p(L/2,y)=0,\quad \partial_y p(x,\pm\, W/2)=0.
    \label{eqn:BCs_curved2}
\end{equation}
We can also expand the boundary conditions of Eq.~(\ref{eqn:BCs_curved2}) in increasing powers of $a$, we find
\begin{equation}
    \begin{cases}
    \tilde\partial_x p^{(0)}(-\tilde L/2,\tilde y)=-\dfrac{v_0\sigma_x}{\kappa},\ p^{(0)}(\tilde L/2,\tilde y)=0, \ \tilde\partial_y p^{(0)}(\tilde x,\pm \,\tilde W/2)=0,\\[0.7em]
    \tilde\partial_x p^{(2)}(-\tilde L/2,\tilde y)=0,\ p^{(2)}(\tilde L/2,\tilde y)=0,\ \tilde\partial_y p^{(2)}(\tilde x,\pm\, \tilde W/2)=0,
    \end{cases}
\end{equation}
with $\tilde L=L/\sigma_x$ and $\tilde W=W/\sigma_x$ (in all this work we implicitly assume that $L> W$). The resolution for $p^{(0)}$ is straightforward $p^{(0)}(\tilde x,\tilde y)=(v_0\sigma_x/\kappa)({\tilde L}/{2}-\tilde x)$,
and the equation for $p^{(2)}$ then takes the form of Eq.~(4) in the main text
\begin{equation}
\begin{cases}
    \Delta p^{(2)}=-(v_0/\kappa)\Delta h\,\partial_x h,\\
    \partial_x p^{(2)}(- L/2, y)=0,\\
    \partial_y p^{(2)}( x,\pm\, W/2)=0,\\
    p^{(2)}( L/2, y)=0,
\end{cases}
\label{eqn:laplace_correction_pressure}
\end{equation}
This set of equations readily suggests an electrostatic analogy.
The pressure problem is equivalent to two-dimensional electrostatics in flat space where $p^{(2)}$ plays the role of the electric potential, and $\lambda = (v_0/\kappa)\Delta h\,\partial_x h$ the role of a surface charge distribution. 
Mass conservation imposes that the spatial average of the charge distribution vanishes. 
It is also worth noting that $\lambda$ derives from a distribution of dipoles $\mathbf p(\mathbf r)$:  $\lambda=-\mathbf{\nabla}\cdot\mathbf{p}$, with
\begin{equation}
\mathbf{p}=\frac{v_0}{2\kappa}\left[(\partial_y h)^2-(\partial_x h)^2\right]\mathbf{\hat x}-\frac{v_0}{\kappa}\partial_x h\,\partial_y h\,\mathbf{\hat y}.
\end{equation}

We can them compute the overall dipole strength and orientation:
\begin{equation}
    \mathbf{P}=\int_{-L/2}^{L/2}\d x\int_{-W/2}^{W/2}\d y\,\mathbf{p}(x,y).
\end{equation}
This expression leads to  Eq.~(5) in the main text for the Gaussian bumps.

It would be tempting to think about the velocity field as the analogue of the electric field that derives from $p$. However the analogy is slightly more subtle as Darcy's law also includes a purely kinematic contribution. 
At second order in $a$
\begin{equation}
\mathbf v^{(2)}=v_0g^{(2)}:\mathbf{e_x}-\kappa \left[\partial_x p^{(2)}\mathbf{e_x}+\partial_y p^{(2)}\mathbf{e_y}\right], 
\label{eqn:v_perturbatif}
\end{equation}
where $\mathbf{e_x}=\hat{\mathbf x}+\partial_x h\,\hat{\mathbf z}$ and $\mathbf{e_y}=\hat{\mathbf y}+\partial_y h\,\hat{\mathbf z}$ are the two tangent vectors to the surface, see also main text.
The first term on the right-hand side corresponds to the projection of the unperturbed flow on the curved surface, while the second term is akin to the electric potential deriving from $p^{(2)}$ and can carry non-local corrections to the flow field. When represented in our 3D Euclidian space the components of the velocity field are given by
\begin{equation}
    \begin{cases}
        v_x=v_0-\left[v_0(\partial_x h)^2+\kappa\partial_x p^{(2)}\right]+O(a^4),\\[0.7em]
        
        v_y=-\left[v_0\partial_x h\partial_y h+\kappa\partial_y p^{(2)}\right]+O(a^4),\\[0.7em]
        v_z=v_0\partial_x h -\left\{v_0\partial_x h\left[(\partial_x h)^2 + (\partial_y h)^2\right]+\kappa\left[\partial_x h\partial_x p^{(2)}+\partial_y h\partial_y p^{(2)}\right]\right\}+ O(a^5).
    \end{cases}
    \label{eqn:velocity_components_perturbation}
\end{equation}

\subsubsection{Multipolar expansion of the pressure field and geometrical screening around isotropic bumps}
Ignoring first the effect of the lateral confinement, {\it i.e.}, away from the boundaries, we can solve Eq.~\eqref{eqn:laplace_correction_pressure}~\cite{jackson}. 
To do so, we introduce the Green's function $G(x,y\vert x',y')$ of Laplace's equation in two dimensions:
\begin{equation}
    p^{(2)}(x,y)=\int_{-L/2}^{L/2}\d x'\int_{-W/2}^{W/2}\d y'\,\lambda(x',y')G(x,y\vert x',y').
    \label{eqn:pgreen}
\end{equation}
In an infinite two-dimensional flat medium $G$ takes the standard form
\begin{equation}
    G( x, y\vert  x', y')=-\frac{1}{4\pi}\ln\left[( x- x')^2+( y - y')^2\right]=-\frac{1}{2\pi}\ln( \rho_>)+\frac{1}{2\pi}\sum_{m=1}^{+\infty}\frac{1}{m}\left(\frac{\rho_<}{\rho_>}\right)^m\cos\left[m(\phi-\phi')\right],
    \label{eqn:green_function_infinite}
\end{equation}
with $ \rho=\sqrt{ x^2+ y^2}$, $\rho'=\sqrt{ x'^2+ y'^2}$, $ \rho_<=\min(\rho,\rho')$, $\rho_>=\max(\rho,\rho')$, $\phi$ the angle between the position vector $\mathbf{r}= x\mathbf{\hat x}+ y\mathbf{\hat y}$ and $\mathbf{\hat x}$, and $\phi'$ the angle between the position vector $\mathbf{r'}= x'\mathbf{\hat x}+ y'\mathbf{\hat y}$ and $\mathbf{\hat x}$. The last equality comes from the relation
\begin{equation}
\sum_{m=1}^{+\infty}\frac{1}{m}r^m\cos(m\theta)=-\frac{1}{2}\ln\left[1-2r\cos(\theta)+r^2\right], \quad \abs{r}<1.
\label{eqn:equation_series_ln}
\end{equation}
Using the series expansion of the Green's function in Eq.~\eqref{eqn:pgreen}, we find in the limit $\rho\gg 1$
\begin{equation}
     p^{(2)}( \rho,\phi)\approx\frac{1}{2\pi}\sum_{m=1}^{+\infty}\frac{1}{m\rho^m}\int_0^{+\infty}\d\rho'\int_0^{2\pi}\d\phi'\,\lambda(\rho',\phi') \left(\rho'\right)^{m+1}\cos\left[m(\phi-\phi')\right],
\end{equation}
which is the so-called multipolar expansion of the electric potential associated with the charge distribution $\lambda(x,y)$. 
We note that the zeroth-order term ($m=0$) which usually scales as  $\ln(\rho)$ vanishes because there is not net electric charge.  
This expansion  is exact if the height function has  finite support (if the coordinates at which the pressure is evaluated are outside the domain where the height function is non-zero), and approximate if the height function decays exponentially.

In the case of isotropic bumps, the equivalent charge distribution is of the form $ \lambda( \rho,\phi)= (v_0/\kappa)\Delta h\,\partial_x h=\Lambda( \rho)\cos(\phi)$. By injecting this expression in the above equation, it is straightforward to show that all terms vanish when integrating over $\phi'$, except for $m=1$, resulting in
\begin{equation}
 p^{(2)}( \rho,\phi)=\frac{\cos(\phi)}{2\rho}\int_0^{+\infty}\d\rho'\Lambda(\rho')(\rho')^2.
\end{equation}
This implies that all multipoles of the equivalent charge distribution for an isotropic bump vanish, except the dipole. We now use that $ \Lambda(\rho)=\partial_\rho h\,\partial_\rho( \rho\partial_\rho h)/ \rho$, and we eventually get
\begin{equation}
     p^{(2)}( \rho,\phi)=\frac{\cos(\phi)}{2\rho}\int_0^{+\infty}\d\rho'\,\rho'\partial_\rho h\,\partial_\rho( \rho'\partial_\rho h)=\frac{\cos(\phi)}{2\rho}\left[\frac{1}{2}\left( \rho'\partial_\rho h\right)^2\right]_0^{+\infty}=0.
\end{equation}
We therefore generalize the result presented in the main text for Gaussian bumps: any axisymmetric curvature heterogeneity is associated with a charge distribution where {\em all} multipoles vanish. 

In the case of anisotropic bumps now, the first term in the multipole expansion is finite, and  for $ \rho\gg 1$ we find
\begin{equation}
    p^{(2)}(\rho,\phi)\approx\frac{1}{2\pi \rho}\int_0^{+\infty}\d\rho'\int_0^{2\pi}\d\phi'\lambda(\rho',\phi')( \rho')^2\cos(\phi-\phi')=\frac{\mathbf{P}\cdot\mathbf{r}}{2\pi \abs{\mathbf{r}}^2},
    \label{eqn:pressure_dipolar_anisotropic}
\end{equation}
where we recognize the definition of the total dipole $\mathbf{P}$ of the equivalent charge distribution. We recover the well-known result of the electric potential created by an electric dipole at the origin in an infinite two-dimensional space.

This essential result tells us that the anisotropy factor is a singular perturbation to the flow field. When $q=1$ flows are screened at a rate given by the decay of $h$, while the flows decays universally as $\sim1/\rho^2$ away from anisotropic bumps.

\subsubsection{Finite-size effects}

In this section, we want to take into account finite-size effects, which are inevitable in any experiment. 
 We can still use our electrostatic analogy. 
 We now need to compute the Green's function for Laplace's equation in a rectangle with similar boundary conditions as in Eq.~(\ref{eqn:laplace_correction_pressure}). 
 In other words, we look for the solution of
\begin{equation}
    \begin{cases}
    \Delta G(x,y\vert x',y')=-\delta(x-x')\delta(y- y'),\\
    \partial_x G(- L/2, y)=0,\\
    \partial_yG( x,\pm\,  W/2)=0,\\
    G( L/2, y)=0.
\end{cases}
\end{equation}

We proceed along the lines of Ref.~\cite{duffy2015green} (which deals with homogeneous Dirichlet boundary conditions). 
We first develop the Green's function in Fourier series with respect to $y$. 
Given our boundary conditions, this development takes the form
\begin{equation}
G(x,y\vert x',y')=\sum_{m=1}^{+\infty}G_m(x\vert x', y')\cos\left(\frac{m\pi( y + W/2)}{W}\right).
\end{equation}
We inject this form into the equation satisfied by $G$, we multiply by $\cos[m\pi( y+ W/2)/W]$, integrate over $ y$, and we find that
\begin{equation}
    \partial_{xx}G_{m}( x\vert x', y')-\left(\frac{m\pi}{W}\right)^2G_{m}( x\vert x', y')=-\frac{2}{W}\cos\left(\frac{m\pi( y'+ W/2)}{ W}\right)\delta( x- x').
    \label{eqn:resolution_green_function}
\end{equation}
We then solve this equation for $ x< x'$ and $ x> x'$ separately and find
\begin{equation}
    G_{m}( x\vert x', y')=\begin{cases}A_m( x', y')\cosh\left(\dfrac{m\pi( x + L/2)}{W}\right)\ \text{if}\  x< x',\\[0.8em]
    B_m( x', y')\sinh\left(\dfrac{m\pi( x - L/2)}{W}\right)\ \text{if}\  x> x'.
    \end{cases}
\end{equation}
The two integration constants are determined by imposing that $G_m$ is continuous for $ x= x'$ while its first derivative has a discontinuity that can be determined by integrating Eq.~(\ref{eqn:resolution_green_function}) with respect to $x$ on a small interval centered around $x'$:
\begin{equation}
    \partial_x G_m( x'^+\vert x', y')-\partial_x G_m( x'^-\vert x', y')=-\frac{2}{W}\cos\left(\frac{m\pi( y'+ W/2)}{ W}\right).
\end{equation}
We eventually end up with
\begin{equation}
    G(x,y\vert x',y')=\frac{2}{\pi}\sum_{m=1}^{+\infty}\dfrac{\displaystyle\cos\left(\frac{m\pi( y + W /2)}{W}\right)\cos\left(\frac{m\pi( y' + W /2)}{W}\right)}{\displaystyle m\cosh\left(\frac{m\pi  L}{ W}\right)}\sinh\left(\frac{m\pi( L/2- x_>)}{ W}\right)\cosh\left(\frac{m\pi( L/2+ x_<)}{W}\right),
\end{equation}
with $ x_>=\max( x, x')$ and $ x_<=\min( x, x')$. This expression is however hardly tractable because this series does not converge uniformly~\cite{melnikov2006computability}. We rearrange the different terms by first expanding the hyperbolic sine term $\sinh\left(m\pi( L/2- x_>)/W\right)=\sinh\left[-m\pi(L/2+ x_>)/W+m\pi L/W\right]$. We get
\begin{equation}
\begin{aligned}
    &G(x,y\vert x',y')=\\
    &\frac{2}{\pi}\sum_{m=1}^{+\infty}\frac{1}{m}\cos\left(\frac{m\pi( y + W /2)}{W}\right)\cos\left(\frac{m\pi( y' + W /2)}{W}\right)\tanh\left(\frac{m\pi L}{W}\right)\cosh\left(\frac{m\pi(L/2+ x_>)}{W}\right)\cosh\left(\frac{m\pi(L/2+ x_<)}{W}\right)\\
    &-\frac{2}{\pi}\sum_{m=1}^{+\infty}\frac{1}{m}\cos\left(\frac{m\pi( y + W /2)}{W}\right)\cos\left(\frac{m\pi( y' + W /2)}{W}\right)\sinh\left(\frac{m\pi(L/2+ x_>)}{W}\right)\cosh\left(\frac{m\pi(L/2+ x_<)}{W}\right),\\
    &=\frac{2}{\pi}\sum_{m=1}^{+\infty}\frac{1}{m}\cos\left(\frac{m\pi( y + W /2)}{W}\right)\cos\left(\frac{m\pi( y' + W /2)}{W}\right)\left[\tanh\left(\frac{m\pi L}{W}\right)-1\right]\\
    &\times\cosh\left(\frac{m\pi(x+L/2)}{W}\right)\cosh\left(\frac{m\pi(x'+L/2)}{W}\right)\\
    &+\frac{2}{\pi}\sum_{m=1}^{+\infty}\frac{1}{m}\cos\left(\frac{m\pi( y + W /2)}{W}\right)\cos\left(\frac{m\pi( y' + W /2)}{W}\right)\cosh\left(\frac{m\pi(x_<+L/2)}{W}\right)\\
    &\times\left[\cosh\left(\frac{m\pi(x_>+L/2)}{W}\right)-\sinh\left(\frac{m\pi(x_>+L/2)}{W}\right)\right].
    \end{aligned}
    \label{eqn:step1_green}
\end{equation}
We can then find an explicit expression for the second sum, after some algebraic manipulations and by using Eq.~(\ref{eqn:equation_series_ln}):
\begin{equation}
\begin{aligned}
    &\frac{2}{\pi}\sum_{m=1}^{+\infty}\frac{1}{m}\cos\left(\frac{m\pi( y + W /2)}{W}\right)\cos\left(\frac{m\pi( y' + W /2)}{W}\right)\cosh\left(\frac{m\pi(x_<+L/2)}{W}\right)\\
    &\times\left[\cosh\left(\frac{m\pi(x_>+L/2)}{W}\right)-\sinh\left(\frac{m\pi(x_>+L/2)}{W}\right)\right]\\
    &=\frac{1}{2\pi}\sum_{m=1}^{+\infty}\frac{1}{m}\left[(-1)^m\cos\left(\frac{m\pi (y+y')}{W}\right)+\cos\left(\frac{m\pi (y-y')}{W}\right)\right]\left[\exp\left(-\frac{m\pi \abs{x-x'}}{W}\right)+e^{-m\pi L/W}\exp\left(-\frac{m\pi (x+x')}{W}\right)\right],\\
    &=-\frac{1}{4\pi}\left\{\ln\left[1+2e^{-\pi\abs{x-x'}/W}\cos\left(\frac{\pi(y+y')}{W}\right)+e^{-2\pi\abs{x-x'}/W}\right]+\ln\left[1-2e^{-\pi\abs{x-x'}/W}\cos\left(\frac{\pi(y-y')}{W}\right)+e^{-2\pi\abs{x-x'}/W}\right]\right.\\
    &\left.+\ln\left[1+2e^{-\pi(x+x'+L)/W}\cos\left(\frac{\pi(y+y')}{W}\right)+e^{-2\pi(x+x'+L)/W}\right]\right.\\
    &\left.+\ln\left[1-2e^{-\pi(x+x'+L)/W}\cos\left(\frac{\pi(y-y')}{W}\right)+e^{-2\pi(x+x'+L)/W}\right]\right\}.
    \end{aligned}
\end{equation}
The first sum in Eq.~\eqref{eqn:step1_green} can also be simplified via other algebraic manipulations:
\begin{equation}
\begin{aligned}
&\frac{2}{\pi}\sum_{m=1}^{+\infty}\frac{1}{m}\cos\left(\frac{m\pi (y+W/2)}{W}\right)\cos\left(\frac{m\pi (y'+W/2)}{W}\right)\left[\tanh\left(\frac{m\pi L}{W}\right)-1\right]\cosh\left(\frac{m\pi (x+L/2)}{W}\right)\cosh\left(\frac{m\pi (x'+L/2)}{W}\right)\\
&=-\frac{1}{\pi}\sum_{m=1}^{+\infty}\frac{\displaystyle \cos\left(\frac{m\pi (y+W/2)}{W}\right)\cos\left(\frac{m\pi (y'+W/2)}{W}\right)}{\displaystyle m\cosh\left(\frac{m\pi L}{W}\right)}e^{-m\pi L / W}\left[\cosh\left(\frac{m\pi (x+x'+L)}{W}\right)+\cosh\left(\frac{m\pi (x-x')}{W}\right)\right],\\
&=-\frac{2}{\pi}\sum_{m=1}^{+\infty}\frac{1}{m}\cos\left(\frac{m\pi (y+W/2)}{W}\right)\cos\left(\frac{m\pi (y'+W/2)}{W}\right)e^{-2m\pi L /W}\cosh\left(\frac{m\pi (x-x')}{W}\right)\\
&+\frac{1}{\pi}\sum_{m=1}^{+\infty}\frac{\displaystyle \cos\left(\frac{m\pi (y+W/2)}{W}\right)\cos\left(\frac{m\pi (y'+W/2)}{W}\right)}{\displaystyle me^{2m\pi L / W}\cosh\left(\frac{m\pi L}{W}\right)}\left[e^{-m\pi L /W}\cosh\left(\frac{m\pi (x-x')}{W}\right)-e^{m\pi L/W}\cosh\left(\frac{m\pi (x+x'+ L)}{W}\right)\right].
\end{aligned}
\end{equation}
The first sum in the above equation can be computed exactly using Eq.~(\ref{eqn:equation_series_ln}), and we eventually derive that
\begin{equation}
    \begin{aligned}
    &G(x,y\vert x',y')=\\
    &-\frac{1}{4\pi}\left\{\ln\left[1+2e^{-\pi\abs{x-x'}/W}\cos\left(\frac{\pi(y+y')}{W}\right)+e^{-2\pi\abs{x-x'}/W}\right]+\ln\left[1-2e^{-\pi\abs{x-x'}/W}\cos\left(\frac{\pi(y-y')}{W}\right)+e^{-2\pi\abs{x-x'}/W}\right]\right.\\
    &\left.+\ln\left[1+2e^{-\pi(x+x'+L)/W}\cos\left(\frac{\pi(y+y')}{W}\right)+e^{-2\pi(x+x'+L)/W}\right]+\ln\left[1-2e^{-\pi(x+x'+L)/W}\cos\left(\frac{\pi(y-y')}{W}\right)+e^{-2\pi(x+x'+L)/W}\right]\right.\\
    &\left.-\ln\left[1+2e^{-\pi(2L-x+x')/W}\cos\left(\frac{\pi(y+y')}{W}\right)+e^{-2\pi(2L-x+x')/W}\right]\right.\\
    &\left.-\ln\left[1-2e^{-\pi(2L-x+x')/W}\cos\left(\frac{\pi(y-y')}{W}\right)+e^{-2\pi(2L-x+x')/W}\right]\right.\\
    &\left.-\ln\left[1+2e^{-\pi(2L+x-x')/W}\cos\left(\frac{\pi(y+y')}{W}\right)+e^{-2\pi(2L+x-x')/W}\right]\right.\\
    &\left.-\ln\left[1-2e^{-\pi(2L+x-x')/W}\cos\left(\frac{\pi(y-y')}{W}\right)+e^{-2\pi(2L+x-x')/W}\right]\right\}\\
    &+\frac{1}{\pi}\sum_{m=1}^{+\infty}\frac{\displaystyle \cos\left(\frac{m\pi (y+W/2)}{W}\right)\cos\left(\frac{m\pi (y'+W/2)}{W}\right)}{\displaystyle me^{2m\pi L / W}\cosh\left(\frac{m\pi L}{W}\right)}\left[e^{-m\pi L /W}\cosh\left(\frac{m\pi (x-x')}{W}\right)-e^{m\pi L/W}\cosh\left(\frac{m\pi (x+x'+ L)}{W}\right)\right].
\end{aligned}
\end{equation}
The remaining sum converges very fast, and few terms (about 5) are enough to obtain a reasonable estimate of the Green's function.

\begin{figure}
    \centering
    \includegraphics[width=0.49\textwidth]{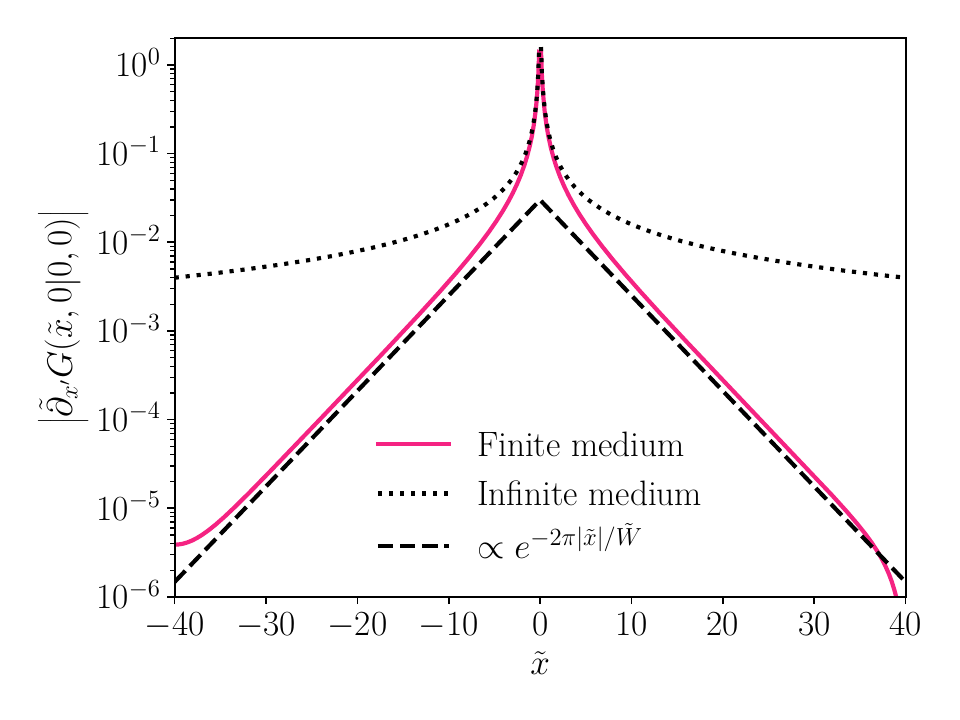}
    \caption{(a) Derivative of the Green's function $\tilde \partial_{x'}G(\tilde x,0\vert 0, 0)$ in rescaled units for Laplace's equation in a rectangular domain of length $\tilde L=80$ and width $\tilde W=76/3$, which represents the equivalent electric potential created by a dipole at the origin. The function crosses over from a power law behavior for $\tilde x \ll \tilde W$ (free space Green's function) to an exponential decay at larger distances.}
    \label{fig:Green}
\end{figure}

\subsubsection{Multipole expansion with finite-size effects}
We can repeat the multipole expansion of Eq.~(\ref{eqn:resolution_green_function}) with the new Green's function which takes into account the boundary conditions. The dipolar contribution now reads
\begin{equation}
    p^{(2)}( x, y)\approx \mathbf{ P}\cdot\mathbf{ \nabla_{\mathbf{ r'}}}G( x, y\vert 0,0),
\end{equation}
where $\mathbf{ \nabla_{\mathbf{r'}}}= \partial_{x'}\mathbf{\hat x}+ \partial_{y'}\mathbf{\hat y}$. For the isotropic bump, $\mathbf{ P}=\mathbf{0}$, and the dipolar contribution still vanishes. For anisotropic Gaussian  bumps, we instead find
\begin{equation}
 p^{(2)}( x, y)\approx \frac{\pi v_0h_0^2}{4\kappa}\left(\frac{\sigma_x}{\sigma_y}-\frac{\sigma_y}{\sigma_x}\right)\partial_{x'}G( x, y\vert0,0).
\end{equation}
In the far-field limit, Eq.~\eqref{eqn:v_perturbatif} then reduces to: 
\begin{equation}
    v_x-1\approx \frac{\pi v_0h_0^2}{4\kappa}\left(\frac{\sigma_y}{\sigma_x}-\frac{\sigma_x}{\sigma_y}\right)\partial_x\partial_{x'}G( x, y\vert0,0)+O(a^4),\quad v_y\approx \frac{\pi v_0h_0^2}{4\kappa}\left(\frac{\sigma_y}{\sigma_x}-\frac{\sigma_x}{\sigma_y}\right)\partial_y\partial_{x'}G( x, y\vert0,0)+O(a^4).
    \label{eqn:components_velocity_green}
\end{equation}
We use Mathematica to compute the function $\partial_{x'}G(x,y\vert0,0)$ and we show its variations in Fig.~\ref{fig:Green} for $y=0$. 
This function displays a clear crossover from a power law behavior for $x \ll W$ (derivative of the free space Green's function) to an exponential decay at larger distances with a characteristic length scale $W/(2\pi)$.

We can go further in the multipole expansion. For  isotropic bumps, we computer the higher-order multipole terms:
\begin{equation}
     p^{(2)}( x, y)=\sum_{k=0}^{+\infty}\sum_{p=0}^k\frac{1}{p!(k-p)!}\partial^{k-p}_{x'}\partial^p_{y'}G( x, y\vert 0,0)\int_{- L/2}^{ L/2}\d x'\int_{- W/2}^{ W/2}\d y'\,\lambda( x', y')( x')^{k-p}( y')^p.
\end{equation}
As mentioned above, $\lambda( x, y)= x\Lambda(\sqrt{ x^2+ y^2})/\sqrt{ x^2+ y^2}$, so that only the terms with $k$ odd and $p$ even remain in the previous development. We thus end up with
\begin{equation}
     p^{(2)}( x, y)=\sum_{k=0}^{+\infty}\sum_{p=0}^k\frac{1}{(2p)!(2k+1-2p)!}\partial^{2k+1-2p}_{x'}\partial^{2p}_{y'}G( x, y\vert 0,0)\int_{- L/2}^{ L/2}\d x'\int_{- W/2}^{ W/2}\d y'\,\lambda( x', y')( x')^{2k+1-2p}( y')^{2p}.
\end{equation}
We now note that $G( x, y\vert x', y')$ can be written as the sum of a function of $ y + y'$ and of another function of $ y- y'$. We thus conclude that $ \partial^2_{y'}G=\partial_y^2G$. Similarly, we note that $ \partial^2_{x'}G=\partial_x^2G$. Using the fact that $ \Delta G=\partial^2_x G+\partial^2_yG=0$ for $ x\neq x'$ and $ y\neq y'$, we  then conclude that
\begin{equation}
     p^{(2)}( x, y)=\sum_{k=0}^{+\infty}\partial^{2k+1}_{x'}G( x, y\vert 0,0)\sum_{p=0}^k\frac{(-1)^p}{(2p)!(2k+1-2p)!}\int_{- L/2}^{ L/2}\d x'\int_{- W/2}^{ W/2}\d y'\,\lambda( x', y')( x')^{2k+1-2p}( y')^{2p}.
\end{equation}
Transforming the integral to polar coordinates and using the fact that $\lambda( \rho,\phi)=\Lambda( \rho)\cos(\phi)$, we obtain 
\begin{equation}
     p^{(2)}( x, y)=\sum_{k=0}^{+\infty}\partial^{2k+1}_{x'}G( x, y\vert 0,0)\int_0^{+\infty}\d\rho'\,\Lambda(\rho')(\rho')^{2k+2}\sum_{p=0}^k\frac{(-1)^p}{(2p)!(2k+1-2p)!}\int_0^{2\pi}\d\phi'\,\cos^{2k+2-2p}(\phi')\sin^{2p}(\phi').
\end{equation}
We have used Mathematica to compute the trigonometric integrals, 
\begin{equation}
    \int_0^{2\pi}\d\phi'\,\cos^{2k+2-2p}(\phi')\sin^{2p}(\phi')=\frac{\pi(2p)!(2k-2p+2)!}{2^{2k+1}p!(k-p+1)!(k+1)!}.
\end{equation}
We can now express the correction to the pressure as
\begin{equation}
     p^{(2)}( x, y)=\sum_{k=0}^{+\infty}\frac{\pi}{4^k(k+1)!}\partial^{2k+1}_{x'}G( x, y\vert 0,0)\int_0^{+\infty}\d\rho'\,\Lambda(\rho')(\rho')^{2k+2}\sum_{p=0}^k\frac{(-1)^p}{p!(k-p)!}.
\end{equation}
We then remark that $\sum_{p=0}^k(-1)^p/[p!(k-p)!]=0^k$, so that only the term for $k=0$ (dipole) does not vanish. Therefore, we conclude that
\begin{equation}
     p^{(2)}( x, y)=\pi\partial_{x'}G( x, y\vert 0,0)\int_0^{+\infty}\d\rho'\Lambda(\rho')(\rho')^2=0,
\end{equation}
as for an infinite medium. Confinement does not alter the symmetry of the pressure perturbations but only their decay in the far-field limit.

\subsection{Numerical resolution and Finite Element Method}
\subsubsection{Methods}
As we could not derive an analytic expression for the pressure field $p(x,y)$ in the case of a general anisotropic bump, or for a finite-size channel, we turn to the numerical resolution of the Laplace-Beltrami equation using a FEM solver provided by the FEniCS package~\cite{langtangen2017solving}. To compare our numerical resolution with our  experiments, we consider a rectangular channel $(x,\ y)\in[-L/2,\ L/2]\times[-W/2,\ W/2]$ with $L=120$~mm and $W=38$~mm and a Gaussian bump. As in our experiments the height field is set to 0 when $|x|>3\sigma_x$ or $|y|>3\sigma_y$ (or when $\rho>3\sigma$ for isotropic bumps).
We express length scales in units of $\sigma_x=1.5$~mm, velocities in units of $v_0$, and pressures in units of $v_0\sigma_x/\kappa$. 
We keep the same notations for the dimensionless and physical quantities. 

We use FEniCS to solve the partial differential equation:
\begin{equation}
\partial_\alpha \left(\sqrt{g}g^{\alpha\beta} \partial_\beta  p\right)=0,\quad 
    \begin{cases}
     \partial_x  p(- L/2, y)=-1,\\
     p( L/2, y)=0,\\
     \partial_y  p(x,\pm\, W/2)=0,
    \end{cases}
    \label{eqn:edp_fenics}
\end{equation}
with $L=80$, and $W= 78/3$ and 
\begin{equation}
    g_{\alpha\beta}=\begin{pmatrix}
        1+(\partial_x h)^2& \partial_x h\,\partial_y h\\
        \partial_x h\,\partial_y h&1+(\partial_y h)^2
    \end{pmatrix}, \quad 
        g^{\alpha\beta}=\frac{1}{g}\begin{pmatrix}1+(\partial_y h)^2&-\partial_x h\,\partial_y h\\
        -\partial_x h\,\partial_y h&1+(\partial_x h)^2,
    \end{pmatrix},\quad g=1+\left[(\partial_x h)^2+(\partial_y h)^2\right].
\end{equation}

Equation~(\ref{eqn:edp_fenics}) is solved thanks to a sparse LU decomposition on a triangular mesh of the rectangle with 250 cells in each direction. We have checked that the results are insensitive to the specific choice of the number of cells.

\subsubsection{Systematic investigation of the influence of the anisotropy factor}

\begin{figure}
    \centering
    \includegraphics[width=0.99\linewidth]{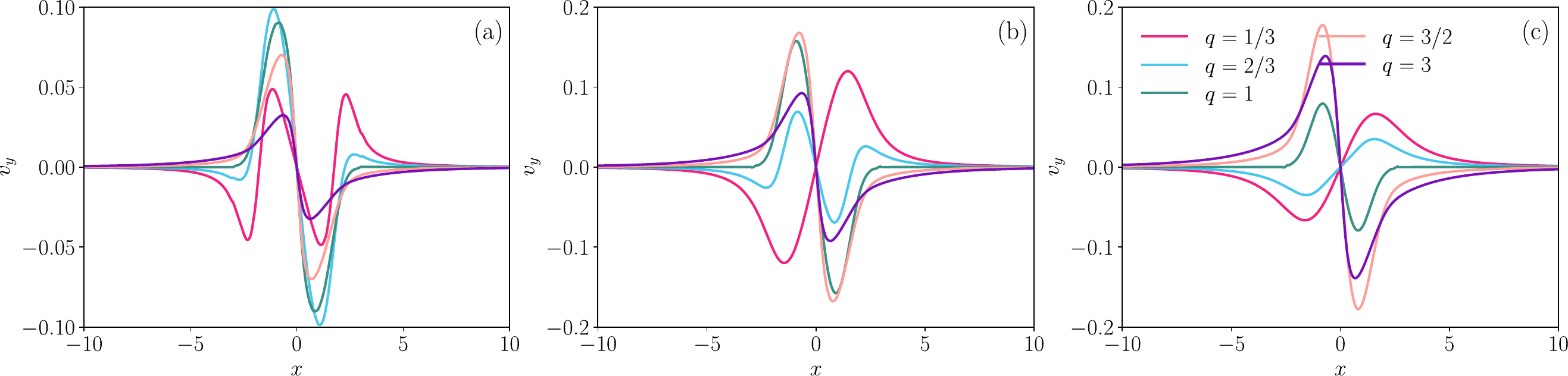}
    \caption{$y$-component of the mean velocity flow for Gaussian bumps with aspect ratio $a=7/3$ and different values of anisotropy $q=\sigma_y/\sigma_x$. The data $ v_y( x, y)$ are obtained via FEM calculations and are represented as a function of the coordinate $ x$ in the direction of the unperturbed flow for several values of $y$ [$ y=0.5,\ 1.5,\ 2.5$ for panels (a), (b), (c) respectively]. For all panels, the legend is similar and is given in panel (c).}
    \label{fig:data_asym_FEM_bis}
\end{figure}

\begin{figure}
    \centering
    \includegraphics[width=0.99\linewidth]{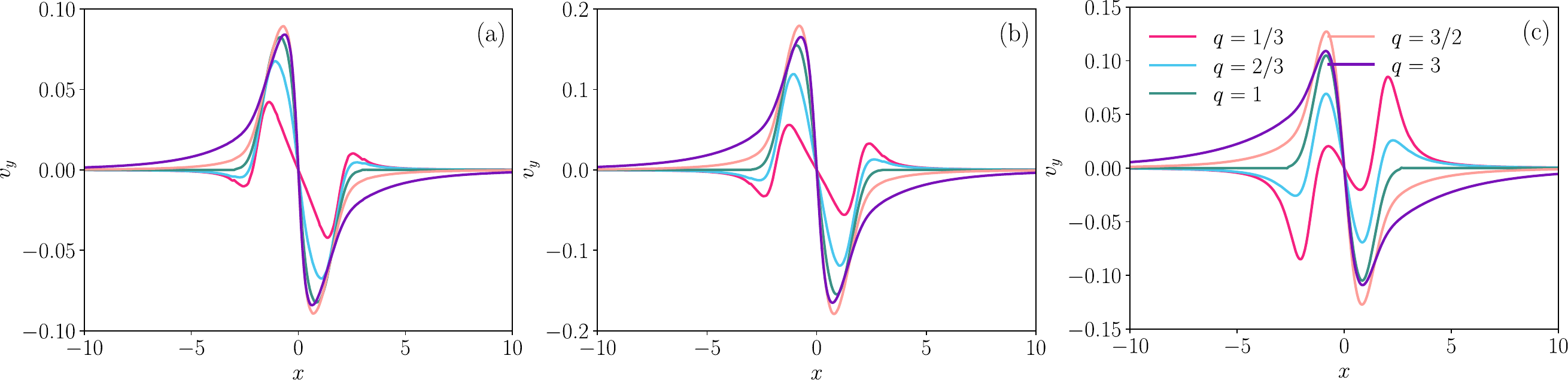}
    \caption{$y$-component of the mean velocity flow for Gaussian bumps with aspect ratio $a=7/3$ and different values of anisotropy $q=\sigma_y/\sigma_x$. The data $ v_y( x, y)$ are obtained via FEM calculations and are represented as a function of the coordinate $ x$ in the direction of the unperturbed flow for several values of $r=y/\sigma_y$ [$r=0.3,\ 0.78,\ 1.5$ for panels (a), (b), (c) respectively]. For all panels, the legend is similar and is given in panel (c).}
    \label{fig:data_asym_FEM}
\end{figure}

\begin{figure}
    \centering
    \includegraphics[width=0.49\linewidth]{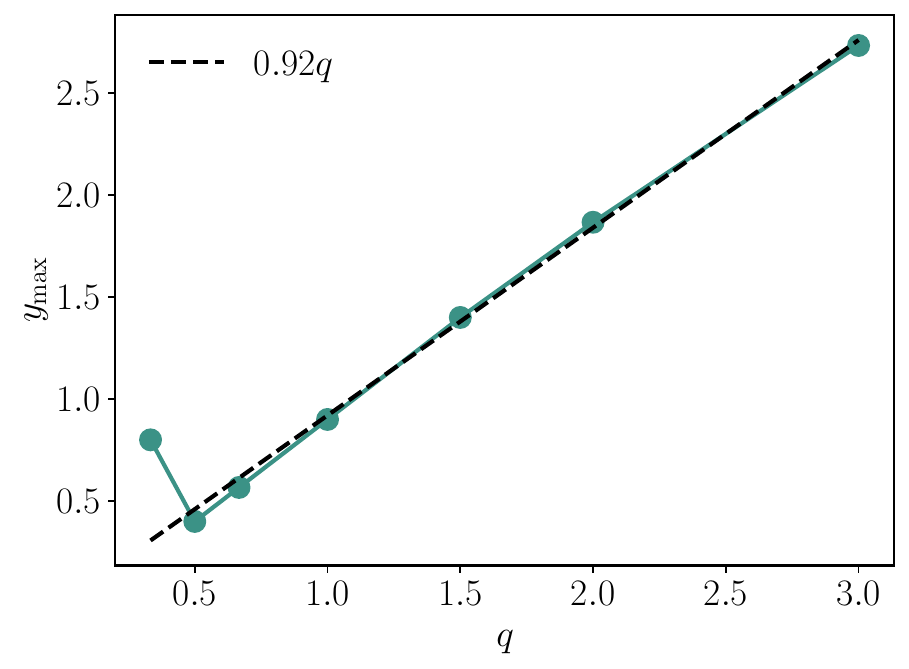}
    \caption{$y$-coordinate $ y_\mathrm{max}$ of the absolute maximum of $ v_y$ for several values of anisotropy $q=\sigma_y/\sigma_x$ of a Gaussian bump of aspect ratio $a=7/3$. These data were obtained via FEM simulations.}
    \label{fig:data_asym_FEM_position_max}
\end{figure}

We use FEM calculations to go beyond experimental findings and systematically study  the influence of the anisotropy of the surface.
In the main text, we focus on the velocity in the $y$-direction. As it vanishes on the $x$-axis ($y=0$), we compute it at a non-zero value of $y$. 
To compare bumps of different anisotropy, we can confront data obtained at fixed $y$ (fixed distance along the $y$-axis in units of $\sigma_x$, see Fig.~\ref{fig:data_asym_FEM_bis}) or at fixed $r=y/q$ (fixed distance along the $y$-axis in units of $\sigma_y$ with $q=\sigma_y/\sigma_x$ the anisotropy of the bump, see Fig.~\ref{fig:data_asym_FEM}). 
We observe that the variation of $v_y$ with $x$ may vary for a given anisotropy, depending on the value of $y$. However, we are mostly interested in the far-field behavior of the velocity field, for which the trends with anisotropy and distance are unchanged for any $ y$ as long as $\abs{ x}\gg 1$. 
We however note that the maximum amplitude of $ v_y$ peaks at $r= y/q\approx 0.78$ (the actual value is $0.92$ for an aspect ratio $a=7/3$ of the bump), see Fig.~\ref{fig:data_asym_FEM_position_max}. This explains the choice of the cut showed in Fig.~3 in the main document.

\begin{figure}
    \centering
    \includegraphics[width=0.95\linewidth]{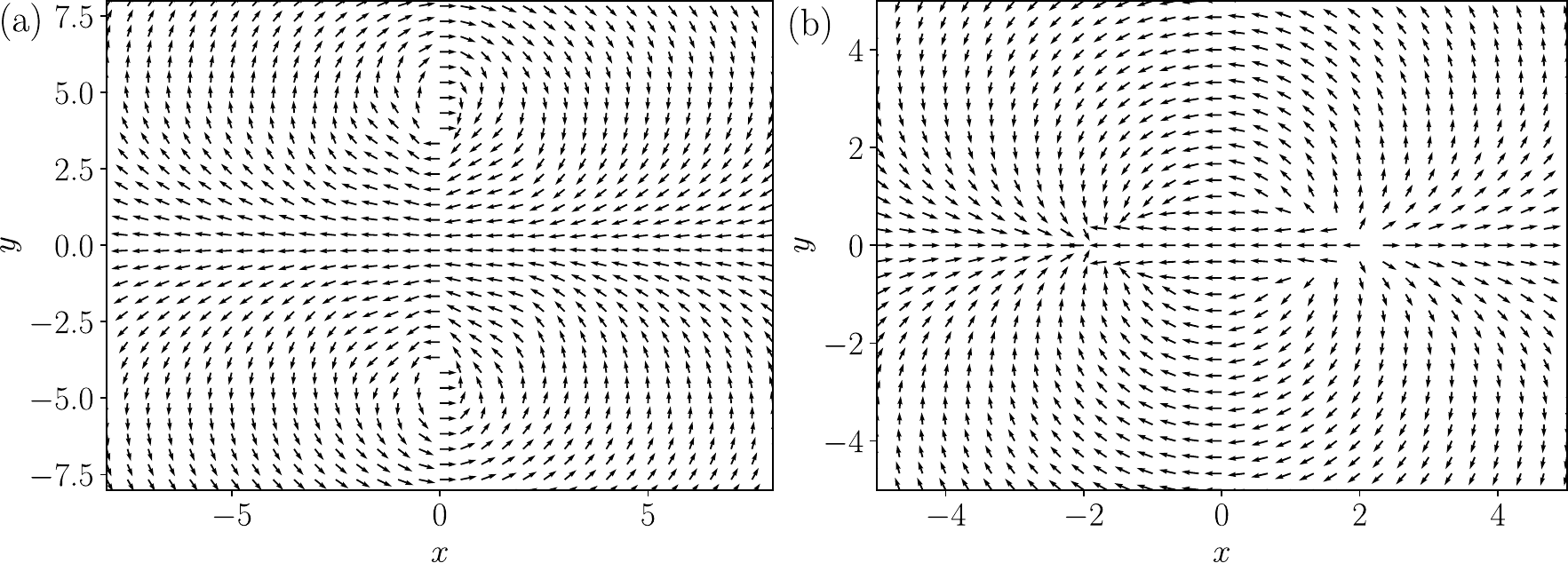}
    \caption{Velocity perturbation ($v_x'$, $v_y' $)=($v_x - 1$, $v_y$) obtained from FEM simulations for Gaussian bumps with aspect ratio $a=7/3$ and two different values of anisotropy $q=\sigma_y/\sigma_x$: (a) $q=3$ and (b) $q=1/3$. All arrows are normalized to be of unit length.}
    \label{fig:data_asym_FEM_map}
\end{figure}

We now analyze more precisely the behavior of the velocity field at large distances, for instance for $r= 0.78$ [see Fig.~\ref{fig:data_asym_FEM}(b)]. 
We observe that the sign of $ v_y$ changes with the anisotropy factor.
$v_y$ has the sign of $ x$ when $q<1$ and reverses its sign  for $q>1$. This is consistent with the electrostatics analogy presented above and in Eq.~(5) in the main text. 
Indeed, if $q>1$, the velocity perturbation induced by the curvature of the channel at large distances from the bump is approximately equivalent to the electric field created by a dipole moment along $-\mathbf{\hat x}$. 
It results in positive (resp. negative) $ v_y$ for negative (resp. positive) $ x$ at sufficiently large distance. 
Conversely, for $q<1$, the equivalent electric dipole moment is along $\mathbf{\hat{x}}$, which results in negative (resp. positive) $ v_y$ for negative (resp. positive) $ x$ at sufficiently large distance. This is further confirmed by a direct inspection of the  $\mathbf v'$ field projected onto the $(x,y)$-plane which shows direct similarities with the electric field created by a point dipole at the origin, see Fig.~\ref{fig:data_asym_FEM_map}. 

\section{Covariant formulation of Navier-Stokes and Darcy's equations}
\subsection{Covariant Navier-Stokes equation}
\label{sec:NS_curved}
In this section, we derive the covariant formulation of the hydrodynamic equations that rule the flow of a fluid in an arbitrary $n$-dimensional curved space. 
For the sake of simplicity, we assume that the $n$-dimensional curved space is embedded in a $(n+1)$-dimensional flat space. 
As a result, the points of the $n$-dimensional curved space are specified by a $(n+1)$-dimensional vector $\mathbf{r}(x^1,\dots,x^n)$ function of $n$ coordinates. 

\subsubsection{Some basics of differential geometry}
We start by recalling few notions of differential geometry that will be useful in the calculations presented below. 
For an introduction to differential geometry in the context of hydrodynamics, the reader is also invited to consult Ref.~[\onlinecite{cai1995Hydrodynamics}].

We first define a basis $(\mathbf{e_1},\dots,\mathbf{e_n})$ of the local tangent plane to the manifold, where $\mathbf{e_\alpha}=\partial_\alpha\mathbf{r}$, with $\partial_\alpha$ standing for the partial derivative with respect to the coordinate $x^\alpha$. We then introduce the $n\times n$ symmetric metric tensor $G_{\alpha\beta}=\mathbf{e_\alpha}\cdot\mathbf{e_\beta}$ [with $\cdot$ the Euclidean scalar product in the $(n+1)$-dimensional embedding flat space] to measure lengths and angles in the curved space. The inverse metric $G^{\alpha\beta}$ is then defined such that $G_{\alpha\gamma}G^{\gamma\beta}=\delta_\alpha^\beta$, where we have used Einstein's convention for the sum over repeated indices, and where $\delta^\alpha_\beta$ is the Kronecker delta. 
The infinitesimal volume is given by $\d V=\sqrt{G}\d x^1\dots\d x^n$, where $\sqrt{G}$ stands for the square root of the determinant of the metric tensor.

We can now represent tangent vectors $\mathbf{v}=v^\alpha \mathbf{e_\alpha}$ in the local basis $(\mathbf{e_1},\dots,\mathbf{e_n})$ using its contravariant components $v^\alpha$. The covariant components of the vector are then $v_\alpha=\mathbf{v}\cdot\mathbf{e_\alpha}=G_{\alpha\beta}v^\beta$. In practice, the metric and its inverse are used to raise or lower the indices. This rule also applies to tensors.

We now need to generalize the notion of derivative to curved manifolds. We thus introduce the covariant derivative $D_\alpha$ which acts on a tensor $T^{\beta\gamma}_{\epsilon\vartheta}$ as follows:
\begin{equation}
    D_\alpha T^{\beta\gamma}_{\epsilon\vartheta}=\partial_\alpha T^{\beta\gamma}_{\epsilon\vartheta}+\Gamma^\beta_{\alpha\iota}T^{\iota\gamma}_{\epsilon\vartheta}+\Gamma^\gamma_{\alpha\iota}T^{\beta\iota}_{\epsilon\vartheta}-\Gamma^\iota_{\alpha\epsilon}T^{\beta\gamma}_{\iota\vartheta}-\Gamma^\iota_{\alpha\vartheta}T^{\beta\gamma}_{\epsilon\iota},
\end{equation}
where $\Gamma^\alpha_{\beta\gamma}$ are the Christoffel symbols characterizing the affine connection. In the following, we use the Levi-Civita connection
\begin{equation}
    \Gamma^\alpha_{\beta\gamma}=\Gamma^\alpha_{\gamma\beta}=\frac{1}{2}G^{\alpha\epsilon}\left(\partial_\beta G_{\epsilon\gamma}+\partial_\gamma G_{\beta\epsilon}-\partial_\epsilon G_{\beta\gamma}\right),
\end{equation}
which is the unique torsion-free connection which is compatible with the metric ($D_\alpha G_{\beta\gamma}=0$). It is straightforward to prove that the covariant derivative has the usual properties of differential operators, for instance regarding the derivative of the product of two tensors. Be careful that the covariant derivatives do not commute because of the curvature of the space. More precisely, for any vector field $v^\alpha$, one has
\begin{equation}
D_\alpha D_\beta v^\gamma-D_\beta D_\alpha v^\gamma=R^\gamma_{\epsilon\alpha\beta}v^\epsilon,
\label{eqn:curvature_commutation}
\end{equation}
where
\begin{equation}
    R^\epsilon_{\gamma\alpha\beta}=\partial_\alpha\Gamma^\epsilon_{\beta\gamma}-\partial_\beta\Gamma^\epsilon_{\alpha\gamma}+\Gamma^\vartheta_{\beta\gamma}\Gamma^\epsilon_{\alpha\vartheta}-\Gamma^\vartheta_{\alpha\gamma}\Gamma^\epsilon_{\beta\vartheta}
        \label{eqn:curv_riemann}
\end{equation}
is the Riemann curvature tensor.

We are now in a position to derive the equations characterizing the flow of a viscous liquid in a curved space in the following sections.

\subsubsection{Mass conservation}
We consider an arbitrary control volume $\mathcal{V}$ of the $n$-dimensional curved space~\cite{henle_hydrodynamics_2010} in order to derive mass conservation equation. Correspondingly, the coordinates $\vec{x}=(x^1,\dots,x^n)$ lie in a set $U_\mathcal{V}$ of $\mathbb{R}^n$. The mass enclosed in the volume at time $t$ reads
\begin{equation}
    M(t)=\iiint_{\vec{x}\in U_\mathcal{V}}\d^n \vec{x}\sqrt{G(\vec{x})}\mu(\vec{x},t),
\end{equation}
with $\mu(\vec{x},t)$ the density field, and where $\d^n\vec{x}$ is a short notation for $\d x^1\dots\d x^n$. Because of mass conservation, the mass enclosed in the volume can only vary because of exchanges with the exterior of the volume, resulting in
\begin{equation}
    \ddroit{M}{t}(t)=-\oiint_{\vec{y}\in U_{\partial\mathcal{V}}}\d^{n-1}\vec{y}\sqrt{g(\vec{y})}n_\alpha(\vec{y}) j^\alpha(\vec{y},t),
\end{equation}
with $\partial\mathcal{V}$ the frontier of the volume $\mathcal{V}$, $\vec{y}\in U_{\partial\mathcal{V}}$ the coordinates parametrizing the points belonging to $\partial\mathcal{V}$, $g_{\alpha\beta}$ the induced metric (pull-back) on $\partial\mathcal{V}$, $g$ its determinant, and $n_\alpha$ the components of the unit vector normal to $\partial \mathcal{V}$ and pointing outwards. In addition $j^\alpha$ stands for the components of the mass current. The divergence theorem then leads to~\cite{cai1995Hydrodynamics}
\begin{equation}
    \partial_t\mu+D_\alpha j^\alpha=0.
\end{equation}

The mass current has the simple expression $j^\alpha=\mu v^\alpha$~\cite{landau2013fluid}, with $v^\alpha$ the components of the velocity field, so that, eventually, mass conservation has the simple form
\begin{equation}
    \partial_t\mu+D_\alpha \left(\mu v^\alpha\right)=0.
    \label{eqn:mass_conservation_cov}
\end{equation}

\subsubsection{Momentum conservation}
To obtain the equation of motion representing momentum conservation, we start from the equations of motion of a dissipative fluid in general relativity~\cite{landau2013fluid, eckart1940thermodynamics}. For this purpose, we consider the $(n+1)\times(n+1)$ space-time metric $\tilde G_{ab}$, such that $\tilde G_{00}=-1$, $\tilde G_{0\alpha}=0$ and $\tilde G_{\alpha\beta}=G_{\alpha\beta}$ with $x^0=ct$. Roman indices range from 0 to $n$ while Greek indices only range from 1 to $n$ and represent spatial indices. For this metric, it is straightforward to show that the Christoffel symbols vanish when they involve at least one time index. 

If we denote with $\tilde D_a$ the covariant derivative associated with this new metric, the equations of motion of the fluid simply read
\begin{equation}
    \tilde D_b T^{ab}=0,
    \label{eqn:momentum_conservation_start}
\end{equation}
where $T^{ab}$ is the momentum-energy tensor. 
When the temperature is constant it reduces to:
\begin{equation}
    T^{ab}=\varepsilon u^a u^b + p\Delta^{ab}-\sigma^{ab}.
    \label{eqn:momentum_energy_tensor1}
\end{equation}
The first two terms correspond to the momentum-energy tensor of a perfect fluid, with $\varepsilon$ its total energy (including its mass energy, kinetic energy and thermal energy) and $p$ the pressure. In addition, we have introduced $\Delta^{ab}=u^au^b+\tilde G^{ab}$ and $u^a=(\gamma, \ \gamma v^\alpha/c)$ the rescaled velocity with $\gamma$ the Lorentz factor. In particular, we note that $\tilde G_{ab}u^au^b=-1$. 
The last term corresponds to the viscous stress tensor, and its first order approximation in gradients of the velocity field reads~\cite{kovtun2012lectures}
\begin{equation}
    \sigma^{ab}=c\eta \tilde G_{de}\left(\Delta^{ac}\Delta^{bd}+\Delta^{ad}\Delta^{bc}-\frac{2}{3}\Delta^{ab}\Delta^{cd}\right)\tilde D_c u^e+c\zeta \tilde D_c u^c \Delta^{ab}=\sigma^{ab}_{(\eta)}+c\zeta \tilde D_c u^c \Delta^{ab},
    \label{eqn:momentum_energy_tensor2}
\end{equation}
where $\eta$ and $\zeta$ are the shear and bulk viscosities respectively.

To derive the fluid's equations of motion, it is convenient to project Eq.~(\ref{eqn:momentum_conservation_start}) on the direction of the velocity and in the orthogonal space. By first projecting in the direction of the velocity, we obtain
\begin{equation}
    -u^b\tilde D_b\varepsilon-(\varepsilon+p)\tilde D_b u^b-\tilde G_{dc}u^c\tilde D_b\sigma^{db}_{(\eta)}+c\zeta\left(\tilde D_b u^b\right)^2=0,
    \label{eqn:momentum_energy_conservation_long}
\end{equation}
where we have used that $\tilde G_{ab}u^a\Delta^{bc}=0$, along with the normalization of the velocity. Then, by taking Eq.~(\ref{eqn:momentum_conservation_start}) and adding Eq.~(\ref{eqn:momentum_energy_conservation_long}) multiplied by $u^a$, we get
\begin{equation}
    \left(\varepsilon+p-c\zeta\tilde D_c u^c\right)u^b\tilde D_b u^a-\tilde D_b \sigma^{ab}_{(\eta)}-\tilde G_{cd}u^au^c\tilde D_b\sigma^{bd}_{(\eta)}+\Delta^{ab}\tilde D_b p-c\zeta \Delta^{ab}\tilde D_b\tilde D_cu^c=0.
    \label{eqn:momentum_energy_conservation_trans}
\end{equation}
We now consider the spatial components of the previous equation $a=\alpha$ in the non-relativistic case, namely, $u^0=1$ and $u^\alpha=v^\alpha/c\ll 1$. As $\varepsilon$ contains the rest energy of the fluid, $\varepsilon\gg p,c\zeta\vert\tilde D_cu^c\vert$, and $\varepsilon\approx \mu c^2$. In this non-relativistic limit, Eq.~(\ref{eqn:momentum_energy_conservation_trans}) becomes
\begin{equation}
    \mu \left(\partial_t v^\alpha+ v^\beta D_\beta v^\alpha\right)=-G^{\alpha\beta}D_\beta p+\tilde D_b\sigma^{\alpha b}_{(\eta)}+\zeta G^{\alpha\beta}D_\beta D_\gamma v^\gamma.  
\end{equation}
We focus here on non-relativistic viscous fluids 
where $\sigma^{\alpha0}=0$ and
\begin{equation}
    \sigma^{\alpha\beta}_{(\eta)}=\eta\left(G^{\alpha\gamma}D_\gamma v^\beta+G^{\beta\gamma}D_\gamma v^\alpha-\frac{2}{3}G^{\alpha\beta}D_\gamma v^\gamma\right),
    \label{eqn:viscous_stress_tensor}
\end{equation}
resulting in
\begin{equation}
    \mu \left(\partial_t v^\alpha+ v^\beta D_\beta v^\alpha\right)=-G^{\alpha\beta}D_\beta p+\eta G^{\beta\gamma}D_\beta D_\gamma v^\alpha+\eta G^{\alpha\beta}R_{\beta\gamma}v^\gamma+\left(\zeta+\frac{\eta}{3}\right) G^{\alpha\beta}D_\beta D_\gamma v^\gamma.
    \label{eqn:NS_curved}
\end{equation}
To derive the above relation, we have used the commutation relation between the covariant derivatives given by Eq.~(\ref{eqn:curvature_commutation}), and we have introduced the Ricci curvature tensor $R_{\alpha\beta}=R^\gamma_{\alpha\gamma\beta}$. Equation~(\ref{eqn:NS_curved}) represents the natural generalization of Navier-Stokes equation to curved manifolds for non-relativistic Newtonian fluid flows. It is important to note that an extra term appears in the right-hand side of the equation, and couples the viscosity to the curvature of the manifold.

In the following we will focus on flows in curved narrow channels, corresponding to incompressible and stationary Stokes flows at sufficiently low Reynolds number $\mathrm{Re}=\mu ev_0/\eta\ll \sigma/e$ (with $v_0$ an order of magnitude of the velocity, and $\sigma\gg e$ a typical distance over which the height function defining the surface varies in the $x$- and $y$-directions). 
In this limit, the governing equations of fluid motion are
\begin{equation}
    \begin{cases}
    D_\alpha v^\alpha = 0, \\
    \eta G^{\beta\gamma}D_\beta D_\gamma v^\alpha+\eta G^{\alpha\beta}R_{\beta\gamma}v^\gamma=G^{\alpha\beta}D_\beta p.
    \label{eqn:Stokes_curved}
    \end{cases}
\end{equation}

\subsection{Generalization of Darcy's law to curved channels}
\label{sec:darcy_curved}
In this section, we show how to derive Darcy's law in three-dimensional channels bounded by two curved two-dimensional surfaces from the constitutive equations of Stokes flow in curved space [see Eq.~(\ref{eqn:Stokes_curved})]. There is no simple solution in the general case, but an approximate solution can be found in the limit where the thickness $e$ of the channel is much smaller than any other length scale involved in the problem (in particular, the typical distance $\sigma$ over which the metric elements change). In this limit, we prove in the following that 

\begin{equation}
        D_\alpha \overline{v^\alpha} =\grandO{He},\quad\overline{v^\alpha} = - g^{\alpha\beta}\partial_\beta p +\grandO{He},
\end{equation}
for $\alpha,\ \beta =x, \ y$, where $\overline{v^\alpha}$ stands for the mean velocity along the normal direction between the two surfaces which bound the curved Hele-Shaw cell, and where $H$ is the local mean curvature of the surface (which is proportional to $\sigma^{-1}$).

\subsubsection{Geometry of the channel}

We first consider a two-dimensional curved surface $\mathcal{S}$ embedded in a flat three-dimensional space. This surface is characterized by its $2\times 2$ metric tensor $g_{\alpha\beta}=\mathbf{e_\alpha}\cdot\mathbf{e_\beta}$, with $\mathbf{e_\alpha}=\partial_\alpha\mathbf{r}$ and $\alpha = x, \ y$. The coordinates' ranges are $x\in[-L/2,\ L/2]$ and $y\in[-W/2,\ W/2]$. If we denote $w\in[0,\ e]$ the coordinate along the normal direction to the surface, the position $\mathbf{R}(x,y,w)$ of any point in the channel satisfies
\begin{equation}
\drond{\mathbf{R}}{w}=\mathbf{N}_w(\mathbf{R}),
\label{eqn:geometry_channel}
\end{equation}
with $\mathbf{R}(x,y,0)=\mathbf{r}(x,y)$ and $\mathbf{N}_w$ the normal unit vector to the surface $\mathcal{S}(w)$ obtained by translation of $\mathcal{S}$ in the local normal direction by an amount $w$.

Our strategy is now to develop Eq.~(\ref{eqn:Stokes_curved}) in powers of $w$ and then to keep only the leading term in the limit $w\to 0$. This requires to expand the metric tensor $G_{\mu\nu}=\partial_\mu\mathbf{R}\cdot\partial_\nu\mathbf{R}$ (with $\mu, \ \nu=x, \ y,\ w$) up to second order in $w$, as the different geometric tensors require at most two consecutive derivatives with respect to the coordinates. As the metric is obtained from the differential element of $\mathbf{R}$, we have to expand the position up to third order in $w$. For this purpose, we proceed order by order.

At first order in $w$, Eq.~(\ref{eqn:geometry_channel}) reads 
\begin{equation}
\mathbf{R}^{(1)}(x,y,w)=\mathbf{r}(x,y)+w\mathbf{n}(x,y)+\grandO{w^2},
\end{equation}
where $\mathbf{n}=(\mathbf{e_x}\times\mathbf{e_y})/\norme{\mathbf{e_x}\times\mathbf{e_y}}$ is the local unit normal vector to the surface $\mathcal{S}(w=0)$ (with $\times$ the Euclidean cross product in the embedding three-dimensional flat space). At second order in $w$, one gets 
\begin{equation}
\mathbf{R}^{(2)}(x,y,w)=\mathbf{r}(x,y)+w\mathbf{n}(x,y)+w^2\mathbf{A}(x,y)+\grandO{w^3}, 
\end{equation}
where $\mathbf{A}$ is an unknown vector. From the above equation and from Eq.~(\ref{eqn:geometry_channel}), one can derive the first correction to the unit normal vector 
\begin{equation}
\mathbf{N}_w^{(1)}(x,y,w)=\mathbf{n}(x,y)+2w\mathbf{A}(x,y)+\grandO{w^2}. 
\end{equation}
On the one hand, imposing that $\mathbf{N}_w^{(1)}$ is normal to $\mathcal{S}(w)$ yields 
\begin{equation}
\d\mathbf{R}\cdot\mathbf{N}_w^{(1)}=\mathbf{n}\cdot\d\mathbf{r}+w\left[\d\mathbf{n}\cdot\mathbf{n}+2\mathbf{A}\cdot\d\mathbf{r}\right]+\grandO{w^2}.
\end{equation}
By definition of the unit normal vector $\mathbf{n}$, $\mathbf{n}\cdot\d\mathbf{r}=0$ and $\d\mathbf{n}\cdot\mathbf{n}=0$, resulting in $\mathbf{A}\cdot\d\mathbf{r}=0$. The last equality is true for any $\d\mathbf{r}$, we thus conclude that $\mathbf{A}$ is proportional to $\mathbf{n}$. On the other hand, the normalization of the normal vector imposes that $\mathbf{N}_w^{(1)}(x,y,w)\cdot \mathbf{N}_w^{(1)}(x,y,w)=1$, and thus that $\mathbf{A}\cdot\mathbf{n}=0$. As $\mathbf{A}$ is parallel and orthogonal to $\mathbf{n}$, we eventually get that $\mathbf{A}=\mathbf{0}$. The same analysis can be conducted at third order in $w$, with the same conclusion that the correction vanishes. As a consequence, we find
\begin{equation}
\mathbf{R}(x,y,w)= \mathbf{r}(x,y)+w\mathbf{n}(x,y)+\grandO{w^4},
\end{equation} 
and the $3\times 3$ metric in the channel reads
\begin{equation}
    G_{\mu\nu}=\left(\begin{NiceArray}{cw{c}{0.5cm}|c}[margin] \Block[c]{2-2}{\widehat g_{\alpha\beta}} & & 0 \\
& & 0\\
\hline
0 & 0 & 1
\end{NiceArray}\right) +\grandO{w^3},
\end{equation}
where
\begin{equation}
    \widehat g_{\alpha\beta}=g_{\alpha\beta}-2wL_{\alpha\beta}+w^2\partial_\alpha\mathbf{n}\cdot\partial_\beta\mathbf{n},
    \label{eqn:new_metric}
\end{equation}
with $L_{\alpha\beta}=-\mathbf{e_\alpha}\cdot\partial_\beta\mathbf{n}$ the symmetric shape operator whose eigenvalues are the local principal curvatures of the surface. In the following, the indices $\alpha,\ \beta,\ \gamma, \ \dots$ equal $x,\ y$ while the indices $\mu, \ \nu,\ \tau,\ \dots$ equal $x,\ y,\ w$. Its inverse alors reads
\begin{equation}
    G^{\mu\nu}=\left(\begin{NiceArray}{cw{c}{0.5cm}|c}[margin] \Block[c]{2-2}{\widehat g^{\alpha\beta}} & & 0 \\
& & 0\\
\hline
0 & 0 & 1
\end{NiceArray}\right) +\grandO{w^3},
\end{equation}
where
\begin{equation}
    \widehat g^{\alpha\beta}=g^{\alpha\beta}+2wg^{\alpha\gamma}g^{\beta\epsilon}L_{\gamma\epsilon}+w^2\left[g^{\alpha\gamma}g^{\beta\epsilon}g^{\vartheta\iota}L_{\gamma\vartheta}L_{\epsilon\iota}-g^{\alpha\gamma}g^{\beta\epsilon}\partial_\gamma\mathbf{n}\cdot\partial_\epsilon\mathbf{n}\right].
\end{equation}
From the metric tensor and its inverse, one easily finds the Christoffel symbols at first order in $w$ (where the symbols with a hat are associated with the three-dimensional metric $G_{\mu\nu}$ while the others are associated with the two-dimensional metric $g_{\alpha\beta}$):
\begin{equation}
\begin{cases}
    \widehat \Gamma^w_{\alpha\beta}=L_{\alpha\beta}-w\partial_\alpha\mathbf{n}\cdot\partial_\beta\mathbf{n}+\grandO{w^2},\\
    \widehat \Gamma^w_{w\mu}=\grandO{w^2},\\
    \widehat \Gamma^\alpha_{\beta\gamma}=\Gamma^\alpha_{\beta\gamma}-wg^{\alpha\epsilon}\left[D_\gamma L_{\beta\epsilon}+D_\beta L_{\gamma\epsilon}-D_\epsilon L_{\beta\gamma}\right]+\grandO{w^2},\\
    \widehat \Gamma^\alpha_{w\beta}=-g^{\alpha\gamma}L_{\beta\gamma}+w \left[g^{\alpha\gamma}\partial_\beta\mathbf{n}\cdot\partial_\gamma\mathbf{n}-2g^{\alpha\epsilon}g^{\gamma\vartheta}L_{\epsilon\vartheta}L_{\beta\gamma}\right]+\grandO{w^2},\\
    \widehat \Gamma^\alpha_{ww}=\grandO{w^2}.
\end{cases}
\end{equation}
Similarly, one can compute the Ricci curvatuve and truncate the calculation at the zeroth order in $w$:
\begin{equation}
    \begin{cases}
    \widehat R_{\alpha\beta}=R_{\alpha\beta}+g^{\gamma\epsilon}\left[L_{\alpha\epsilon}L_{\beta\gamma}+L_{\alpha\gamma}L_{\beta\epsilon}-L_{\alpha\beta}L_{\gamma\epsilon}\right]-\partial_\alpha\mathbf{n}\cdot\partial_\beta\mathbf{n}+\grandO{w},\\
    \widehat R_{w\alpha}=g^{\beta\gamma}\left[D_\alpha L_{\beta\gamma}-D_\beta L_{\alpha\gamma}\right]+\grandO{w},\\
    \widehat R_{ww}=g^{\alpha\beta}\left[g^{\gamma\epsilon}L_{\alpha\gamma}L_{\beta\epsilon}-\partial_\alpha\mathbf{n}\cdot\partial_\beta\mathbf{n}\right]+\grandO{w}.
    \end{cases}
    \label{eqn:ricci}
\end{equation}

\subsubsection{Expansion of Stokes flow in the limit of a narrow channel}

We start from the equation of mass conservation which reads, at leading order in $w$,
\begin{equation}
    \partial_\alpha v^\alpha+\partial_w v^w+\Gamma^\alpha_{\alpha\beta}v^\beta=0.
    \label{eqn:mass_conservation_approx}
\end{equation}
We now use the above equation to obtain an estimate of the different components of the velocity, and we obtain that $v^x\sim v^y$ and $v^w\sim v^xe/\sigma\ll v^x$, because $e\ll \sigma$. Terms at next-to-leading order in the above equation are typically of order $He$, where $H$ is the local mean curvature of the surface defined as $2H=g^{\alpha\beta}L_{\alpha\beta}$. 


We now analyse the equations describing momentum conservation. For the components in the tangent plane to the surface, the conservation law reads
\begin{equation}
    \eta G^{\mu\nu} D_\mu D_\nu v^\alpha+\eta \widehat g^{\alpha\beta}\widehat R_{\beta\mu}v^\mu=\widehat g^{\alpha\beta}\partial_\beta p.
\end{equation}
The first term can be computed after standard manipulations:
\begin{equation}
    G^{\mu\nu} D_\mu D_\nu v^\lambda=G^{\mu\nu}\partial_{\mu\nu}v^\lambda+G^{\mu\nu}\partial_\mu\left(\widehat\Gamma^\lambda_{\nu\tau}v^\tau\right)+G^{\mu\nu}\widehat\Gamma^\lambda_{\mu\tau}\left(\partial_\nu v^\tau+\widehat\Gamma^\tau_{\nu\varpi}v^\varpi\right)-G^{\mu\nu}\widehat\Gamma^\tau_{\mu\nu}\left(\partial_\tau v^\lambda+\widehat\Gamma^\lambda_{\tau\varpi}v^\varpi\right).
\end{equation}
Using the fact that $e\ll \sigma$ with the above equations and Eq.~(\ref{eqn:ricci}), we find that the momentum conservation equation reduces to
\begin{equation}
    \eta \partial_{ww}v^\alpha=g^{\alpha\beta}\partial_\beta p.
    \label{eqn:momentum_conservation_Darcy}
\end{equation}
for $\alpha=x, \ y$, at leading order in $w/\sigma$. Once again, terms at next-to-leading order are typically of order $He$.

We now perform the same analysis for the last component of the momentum conservation equation. At leading order in $w$, it is clear from the expression of the Ricci curvature tensor and the covariant derivatives of the velocity field that the equation reduces to 
\begin{equation}
    \eta \partial_{ww}v^w=\partial_w p.
    \label{eqn:momentum_conservation_Darcy2}
\end{equation}
Comparing Eq.~(\ref{eqn:momentum_conservation_Darcy}) and Eq.~(\ref{eqn:momentum_conservation_Darcy2}) implies that $\partial_w p\ll \partial_\alpha p$, and therefore the above equation simply reads
\begin{equation}
    \partial_w p=0.
    \label{eqn:momentum_conservation_Darcy3}
\end{equation}

In consequence, Stokes flow in a curved channel is governed by Eq.~(\ref{eqn:momentum_conservation_Darcy}) and Eq.~(\ref{eqn:momentum_conservation_Darcy3}) at the lowest order in the channel thickness, which admit a simple solution. The latter can be computed by differentiating Eq.~(\ref{eqn:momentum_conservation_Darcy}) with respect to $w$ and using Eq.~(\ref{eqn:momentum_conservation_Darcy3}) to obtain $\partial_{www} v^\alpha =0$. The velocity profile in the channel is thus analogous to Poiseuille flow. Using the fact that the velocity has to vanish on the two bounding surfaces, we derive that
\begin{equation}
    v^\alpha(x,y,w)=\frac{w(w-e)}{2\eta}g^{\alpha\beta}(x,y)\partial_\beta p(x,y).
    \label{eqn:velocity_channel_curved}
\end{equation}

We now introduce the mean velocity along the normal coordinate $w$:
\begin{equation}
    \overline{v^\alpha}(x,y)=\frac{1}{e}\int_0^e\d w\,v^\alpha(x,y,w)=-\kappa g^{\alpha\beta}(x,y)\partial_\beta p(x,y),
    \label{eqn:darcy_curved}
\end{equation}
where $\kappa=e^2/(12\eta)$ represents the permeability of the curved channel. The above equation corresponds to a linear transport equation as the mean flux is proportional to the opposite of the gradient of the pressure profile. Equation~(\ref{eqn:darcy_curved}) generalizes Darcy's law to curved manifolds, which was already advocated in several references without a proper demonstration~\cite{entov1997viscous, brandao_suppression_2014, brandao2015viscous, brandao_capillary_2017}. 
We mention that the equations of motion for the fluid in the curved Hele-Shaw cell are similar to the ones describing the flow of superfluids on a curved surface, where the pressure plays the role of the local phase of the wavefunction~\cite{turner2010vortices}.

We eventually obtain an equation verified by the pressure field alone by averaging Eq.~(\ref{eqn:mass_conservation_approx}) along the normal direction and using the fact that the velocity has to vanish on the two bounding surfaces:
\begin{equation}
    D_\alpha\left(g^{\alpha\beta}\partial_\beta p\right)=\frac{1}{\sqrt{g}}\partial_\alpha\left(\sqrt{g}g^{\alpha\beta}\partial_\beta p\right)=\Delta_\mathrm{LP}p=0,
\end{equation}
which coincides with Eq.~\eqref{eqn:laplace_curved} given above. This partial differential equation should be complemented by boundary conditions. Here, and consistently with the experiments, we impose the flow at the entrance and the pressure at the end (that we can set to zero without loss of generality) of the channel. Besides, we impose that the normal velocity vanishes on the two extremities of the width of the channel. This results in the following boundary conditions for the pressure field:
\begin{equation}
    g^{x\alpha}\partial_\alpha p(-L/2,y)=-v_0/\kappa, \quad p(L/2,y)=0,\quad g^{y\alpha}\partial_\alpha p(x,\pm\, W/2)=0.
    \label{eqn:BCs_curved}
\end{equation}

\bibliography{./biblio.bib}